\documentclass[twocolumn, nofootinbib, floatfix]{aastex631}

\usepackage{bm}
\usepackage{amsmath}
\usepackage{graphicx}
\usepackage{natbib}

\DeclareMathOperator\sign{sign}

\begin{document}

\title{Determination of Lens Mass Density Profile from Strongly-Lensed Gravitational-Wave Signals}

\author{Mick Wright}
\affiliation{SUPA, School of Physics and Astronomy, University of Glasgow, Scotland}

\author{Justin Janquart}
\affiliation{Department of Physics, Institute for Gravitational and Subatomic Physics (GRASP), Utrecht University, Princetonplein 1, NL-3584 CC Utrecht, The Netherlands}
\affiliation{Nikhef --- National Institute for Subatomic Physics, Science Park, NL-1098 XG Amsterdam, The Netherlands}

\author{Martin Hendry}
\affiliation{SUPA, School of Physics and Astronomy, University of Glasgow, Scotland}

\begin{abstract}
As the interferometers detecting gravitational waves are upgraded, improving their sensitivity, the probability of observing strong lensing increases. Once a detection is made, it will be critical to gain as much information as possible about the lensing object from these observations. In this work, we present a methodology to rapidly perform model selection between differing mass density profiles for strongly lensed gravitational wave signals, using the results of the fast strong lensing analysis pipeline GOLUM\@. We demonstrate the validity of this methodology using some illustrative examples adopting the idealised singular isothermal sphere and point mass lens models. We take several simulated lensed signals, analyse them with GOLUM and subject them to our methodology to recover both the model and its parameters. To demonstrate the methodology's stability, we show how the result varies with the number of samples used for a subset of these injections. In addition to the analysis of simulations, we also apply our methodology to the gravitational wave event pair GW191230--LGW200104, two events with similar frequency evolutions and sky locations, which was analysed in detail as a potential lensing candidate but ultimately discarded when considering the full population and the uncertain nature of the second event. We find a preference for the singular isothermal sphere model over the point mass, though our posteriors are much wider than for the lensed injections, in line with the expectations for a non-lensed event. The methodology developed in this work is made available as part of the \textsc{Gravelamps} package of software. 

\end{abstract}

\section{Introduction}\label{sec:introduction}
In the wake of a rapidly increasing number of gravitational wave (GW) detections, with 90 confirmed detections as of the latest catalog of events, GWTC-3 \citep{gwtc3}, and with the coming years set to deliver significant improvements in the sensitivity of the existing network of ground-based GW detectors, currently comprising Advanced LIGO \citep{advancedligo}, Advanced Virgo \citep{advancedvirgo}, and KAGRA \citep{kagra_overview, kagra_design, kagra_configuration}, the opportunities are manifest for the nascent field of GW astronomy to pursue a wide variety of scientific investigations.

One example of a particularly active area of GW research is that of GW lensing \citep{o2_lensing, o3a_lensing, o3_lensing, o3_technical}. Analogously to the deflection of light by a strong gravitational field~\citep{Einstein_lensing}---which formed one of the first major tests of Einstein's general relativity (GR) \citep{eddington_experiment}---GW signals may be lensed by intervening objects such as stars, black holes, galaxies and galaxy clusters \citep{focusing_gravitational_radiation, wang_lensing, takahashi_nakamura_lensing, smith_cluster_lensing, mishra_microlensing_population}. 

As of the end of third observing run of the current ground-based detector network, searches for lensed GW signals have not yet yielded a confirmed detection \citep{o3a_lensing, o3_lensing, o3_technical}. However, with the increased detection rate of GW events expected during the next observing run, there is a signficant probability that one or more of these detections may be identified as \textit{strongly lensed} \citep{smith_cluster_lensing, smith_cluster_lensing_mnras, ng_rate_galaxy_lensing, li_lensing_rate, wierda_rgal, xu_rate, smith_rate_2023}. In this scenario the GW signal would be lensed by an intervening galaxy \citep{dai_galaxy_lensing, ng_rate_galaxy_lensing, robertson_strong_lensing} or galaxy cluster \citep{smith_cluster_lensing, smith_cluster_lensing_mnras, robertson_strong_lensing}, forming multiple distinct ``images''. These images would be potentially magnified, time shifted, and overall phase shifted copies of the original signal that may be detected as repeated events \citep{dai_lensed_waveforms, golum_paper, liu_selection_effects, hanabi_paper, golum_update_paper}.

Detection of such a lensed event would not only be of enormous scientific interest in itself, but would also immediately leverage additional scientfic benefit from the information carried by lensed GW signals from compact binary coalescence events---much as the detection of GW signals from such events in the first place has opened the door to completely new ways in which to investigate these phenomena. Examples of the kinds of investigations that could be enabled by lensed GW signals include, but are not limited to, improving the localisation of detected GW events \citep{hannuksela_lensing_localisation, Wempe:2022zlk}, precision cosmography \citep{sereno_lensing_cosmography, liao_cosmography}, or tests of the speed of gravity \citep{fan_speed_of_gravity}, or general relativity itself \citep{ezquiaga_testing_gr, goyal_testing_gr}.

Moreover, investigations of lensed GW systems may also reveal import information about the nature of the lens. For example, it has been demonstrated that the mass density profile of the lens will influence the properties of lensed GW signals \citep{takahashi_nakamura_lensing, cao_lensing}, so that observation of the latter may allow the former to be constrained. Investigations have therefore been carried out into how effectively GW observations can allow the characterisation of the signal and thus constraint of the lens model parameters in both the microlensing \citep{mishra_microlensing_population}---where lenses would be inidividual compact objects such as stars or black holes---and strong lensing regimes \citep{antonio_lens_modelling, hannuksela_lensing_localisation, Wempe:2022zlk, tambalo_lenses} regimes for specific models of lens. Previous work has also taken the additional step of performing model selection for microlensed GW signals \citep{gravelamps_paper}. However, it is important to note that within that work, the presented methodology was applicable to only signle images and this proved to be insufficient to distinguish between lensing models when observing the individual images of strongly-lensed GW signals.

In this work, we present a new methodology for rapidly performing model selection on the results of model-agnostic joint parameter estimation analyses of strongly lensed multiplets. We demonstrate the effectiveness of this methodology by using it to analyse 65 simulated pairs of lensing signals and, in addition, investigate the pair GW191230 and LGW200104 which were considered by the LVK lensing search to be the pair of events with the highest likelihood of lensing---albeit we stress that the non-lensed hypothesis was still preferred and that LGW200104 is disfavoured as a real GW event by its extremely low $p_{\textrm{astro}}$~\citep{o3_technical}.

The work is structured as follows. In Section~\ref{sec:strong-lensing}, we introduce the basics of strong lensing including the theoretical basis, some potential models of lens, as well as the means by which strong lensing may be searched for model-agnostically in GW data. Section~\ref{sec:method} then outlines the methodology by which we may perform model selection on the results of this model agnostic analysis. Section~\ref{sec:testing} demonstrates the testing that has been done to verify the validity of the methodology both on the set of injections and on the trigger data outlined above. Finally, we present the conclusions and future outlook from this work in Section~\ref{sec:conclusion}.

\section{Strong Lensing of Gravitational Waves}\label{sec:strong-lensing}
In general, lensing of GWs is an amplification process. A standard, non-lensed GW signal---which is described by the phase amplitude $\phi_{obs}(w, \bm{\eta})$ where $w$ is the dimensionless frequency and $\bm{\eta}$ is the displacement of the source from the optical axis---experiences an \textit{amplification} to become the observed lensed signal, $\phi^{L}_{obs}(w, \bm{\eta})$. The ratio between these wave amplitudes is termed the \textit{amplification factor} ($F$) and is such that the relation between the unlensed ($h(f)$) and lensed ($h^{L}(f)$) GW strains is:

\begin{equation}
	h^{L}(f) = F(f) \times h(f).
	\label{eq:lensing-amplification}
\end{equation}

This amplification factor varies depending upon the mass density profile of the lensing object. However, it can be calculated for any profile using a single general expression \citep{schneider_gravitational_lenses, takahashi_nakamura_lensing}: 

\begin{equation}
	F(w,y) = \frac{w}{2\pi i} \int \mathrm{d}^{2} x \exp \left[ iwT\left(x,y\right)  \right],
	\label{eq:amplification-factor-general}
\end{equation}

where $y$ is a dimensionless form of the displacement from the source ($\bm{\eta}$ as described above) and the function $T(x,y)$ yields the dimensionless time delay.

In the case of strong lensing, the mass of the lensing object is very high (which yields $w\gg1$) and the geometric optics approximation is valid. This results in only the stationary points of the time delay function contributing to the integral, reducing it to a summation over these points \citep{nakamura_review_lensing}. These stationary points correspond to the individual images produced in the strong lensing process and the amplification of the $j^{\textrm{th}}$ such image is given by \citep{takahashi_nakamura_lensing, dai_lensed_waveforms, ezquiaga_phase_effects}

\begin{equation}
	F_j(f) = \sqrt{|\mu_j|} \exp \left[ iwft_{j} - i\pi n_{j} \sign(f) \right].
	\label{eq:amplification-factor-general-geometric}
\end{equation}

As can be seen, this yields three separate observable effects on the lensed signals: the magnification ($\mu_j$) of the signal's amplitude, the time delay ($t_{j}$) between the signals due to the difference in path length between the images, and the Morse phase ($n_{j}$). This final property is an overall phase shift of the waveform that may take one of three distinct values, $n = 0,1/2,1$ depending on whether the image corresponds to a minimum, saddle point, or maximum of the time delay function respectively.

\subsection{Lensing Models}\label{subsec:lensing-models}

As has been mentioned above, the general functional form of the amplification factor is given by Eq.~\eqref{eq:amplification-factor-general}, but its final expression will vary for each different model for the mass density profile of the lens. There are many such models that have been considered to describe possible lenses. Here, we briefly describe the two models that have been used in this work as examples, to demonstrate the validity of the methodology. These models were chosen due to their ability to be solved analytically which allows for direct testing of the method as well as the fact that they produce two and only two images which again allows a complete treatment of these two models as testing cases. In follow-up work we will expand our analysis to consider more general mass density profiles that provide large multiplet image sets, so that the effect of detecting a subset of images may be taken into consideration.

\subsubsection{Point Mass}\label{subsubsec:point-mass}

The simplest model of the lensing object is that of a point mass. Such a lens produces two images, regardless of the lens-source configuration. One image will be a minimum of the time delay function, and one will be at a saddle point. Consequently, following on from Eq.~\eqref{eq:amplification-factor-general-geometric}, the amplification factor is \citep{takahashi_nakamura_lensing}

\begin{equation}
	F_{geo}^{\mathrm{PM}}(w,y) = \left|\mu_{+}\right|^{1/2} - \left|\mu_{-}\right|^{1/2} e^{iw\Delta T}.
	\label{eq:point-mass-amplification}
\end{equation}

The image magnifications for the point mass case are given in terms of the dimensionless source position, $y$, by $\mu_{\pm} = 1/2 \pm (y^2 + 4)/(2y \sqrt{y^2 + 4})$ and the time delay between the two images is given by~$\Delta T = y(\sqrt{y^4 + 2})/2 + \ln{\left((\sqrt{y^2 + 4} + y)/ (\sqrt{y^2 + 4} - y)\right)}$. 

\subsubsection{Singular Isothermal Sphere}\label{subsubsec:sis}

As a step up in complexity from the point mass profile, the Singular Isothermal Sphere (SIS) is widely used to describe the dark matter halos of galaxies, due to its ability to describe the flat rotation curves observed for these systems \citep{binney_tremaine_galactic_dynamics}. The SIS profile models the galaxy as an extended luminous matter object embedded within a larger dark matter halo. However, the weakness of this profile is that it suffers from a non-physical central singularity.

For an SIS lens, the configuration of source and lens may alter the number of images produced, with in one case a single image produced at higher source positions, and in the other case two produced at lower. Consequently, the amplification factor---again following Eq.~\eqref{eq:amplification-factor-general-geometric}---is piecewise and given by \citep{takahashi_nakamura_lensing}
\begin{equation}
	F_{geo}^{\mathrm{SIS}}(w,y) =
	\begin{cases}
		|\mu_{+}|^{1/2} - i|\mu_{-}|^{1/2} e^{iw\Delta T} & \text{if } y < 1\\
		|\mu_{+}|^{1/2}					  & \text{if } y \geq 1
	\end{cases},
	\label{eq:sis-amplification}
\end{equation}

where in this case the magnifications are given by $\mu_{\pm} = \pm 1 + 1/y$ and the time delay is given by $\Delta T = 2y$.

\subsection{Strong Lensing Identification}\label{subsec:strong-lensing-detection}

There are a number of means by which one may identify a strong lensing multiplet.  These range from examining the overlap between the GW source posteriors derived from the individual images, as outlined in~\cite{harris_posterior_overlap}, which is speedy but does not provide confirmation of lensing status---and more importantly in our context does not provide detailed information on the observed lensing parameters---to full joint parameter estimation of the two image candidates using the method described in~\cite{liu_selection_effects, hanabi_paper, golum_update_paper}, which is complete but computationally expensive. In this work, we focus on the methodology presented in~\cite{golum_paper, golum_update_paper} implemented in the \textsc{python} package \textsc{GOLUM}~\citep{git_golum} as a middle ground in which events are provided with sufficiently accurate estimates on the observable lensing parameters, as well as providing confirmation of lensing status. Whilst we refer the reader to~\cite{golum_paper, golum_update_paper} for a complete explanation of how this identification is performed, we briefly outline the methodology here---for simplicity considering the two-image case only, although the method is valid for any number of images.

A pair of lensed images, $h_{L}^{j}\left(t_{j};\bm{\theta},\bm{\Lambda}_{j}\right)$, are described in terms of the binary parameters, $\bm{\theta}$ and the individual image parameters, $\bm{\Lambda}_{j}$ for the $j^{\mathrm{th}}$ image. When examined under the lensing hypothesis, $\mathcal{H}_{L}$, under which the binary parameters should be identical, the joint evidence neglecting selection effects is given by~\citep{liu_selection_effects, hanabi_paper}
\begin{equation}
	\begin{aligned}
		p\left(d_{1}, d_{2}|\mathcal{H}_{L}\right) = \int p\left(d_{1}|\bm{\theta}, \bm{\Lambda}_{1}\right)
								  p\left(d_{2}|\bm{\theta}, \bm{\Lambda}_{2}\right) \\
		\times p\left(\bm{\theta}, \bm{\Lambda}_{1}, \bm{\Lambda}_{2}\right)
		\mathrm{d}\theta \mathrm{d}\bm{\Lambda}_{1} \mathrm{d}\bm{\Lambda}_{2}
	\end{aligned},
	\label{eq:joint-evidence}
\end{equation}

where the term $p\left(\bm{\theta}, \bm{\Lambda}_{1}, \bm{\Lambda}_{2}\right)$ is the prior, and the terms $p\left(d_{j}|\bm{\theta}, \bm{\Lambda}_{j}\right)$ are the individual likelihoods~\citep{veitch_vechhio_likelihood}. This may be compared with the joint evidence in the unlensed hypothesis, $\mathcal{H}_{U}$, which is the product of the individual likelihoods: $p(d_1, d_2 | \mathcal{H}_U)  = p(d_1 | \mathcal{H}_U)p(d_2 | \mathcal{H}_U)$. The two are compared in the ``\textit{coherence ratio}'':
\begin{equation} 
\mathcal{C}^{L}_{U} = \frac{p(d_1, d_2 | \mathcal{H}_L)}{p(d_1 | \mathcal{H}_U)p(d_2 | \mathcal{H}_U)} \, .
\end{equation}
Calculation of this joint evidence and coherence ratio in full is the basis for complete joint parameter estimation such as that outlined in~\cite{liu_selection_effects, hanabi_paper}. 

However, the process of calculating the coherence ratio may be sped up by instead considering the \textit{conditional} evidence alongside the individual evidences. This allows acceleration of computation by means of importance sampling and a look-up table---the full details of which are described in~\cite{golum_paper, golum_update_paper}. To outline how this is done, the joint evidence is rewritten in terms of the conditional evidence as
\begin{equation}
	p\left(d_{1}, d_{2}|\mathcal{H}_{L}\right) = p\left(d_{1}|\mathcal{H}_{L}\right)
						     p\left(d_{2}|d_{1},\mathcal{H}_{L}\right). 	
	\label{eq:conditioned-evidence}
\end{equation}

The conditional evidence $p\left(d_{2}|d_{1},\mathcal{H}_{L}\right)$ may be evaluated as a ``marginalised'' likelihood of the form
\begin{equation}
	p\left(d_{2}|d_{1},\mathcal{H}_{L}\right) = \int \left<p\left(d_{2}|\bm{\Theta},\bm{\Phi}\right)\right>
						    _{p\left(\bm{\Theta}|d_{1}\right)}
						   p(\bm{\Phi})\mathrm{d}\bm{\Phi}
	\label{eq:conditioned-evidence-integral}
\end{equation}

where the binary parameters, $\bm{\theta}$, have been replaced with the effective parameters, $\bm{\Theta}$ which absorb the lensing magnification into the observed luminosity distance ($D_L^{\mathrm{obs}} = D_L/\sqrt{\mu}$, where $D_L$ is the source luminosity distance) and the time delay into the observed coalescence time ($t_c^{\mathrm{obs}} = t_c + t$, where $t_c$ is the coalescence time without lensing and $t$ is the time delay),  and the individual image parameters $\bm{\Lambda}_{j}$ have been replaced with the relative image parameters $\bm{\Phi}$, i.e.\ the parameters relative to the first of the images. 

The first term of the integrand is simply the likelihood of the second event averaged over the posterior samples of the first event. Whilst this term would already be faster to compute than the full joint evidence from Eq.~\eqref{eq:joint-evidence}, it is further sped up as the reuse of the posterior samples from the first event allows the construction of a lookup table allowing speedy computation of Eq.~\eqref{eq:conditioned-evidence-integral}.

This culminates in the full calculation of the coherence ratio as
\begin{equation}
	\mathcal{C}^{L}_{U} = \frac{p\left(d_{1}|\mathcal{H}_{L}\right)}
				   {p\left(d_{1}|\mathcal{H}_{U}\right)}
			      \frac{p\left(d_{2}|d_{1},\mathcal{H}_{L}\right)}
				   {p\left(d_{2}|\mathcal{H}_{U}\right)}.
	\label{eq:coherence-ratio-conditioned}
\end{equation}

We note that this coherence ratio is not a full Bayes factor as it does not include the selection effects outlined in~\cite{hanabi_paper}. Nevertheless, their inclusion can be done straightforwardly as a post-processing step to the \textsc{GOLUM} analysis.

\section{Model Selection of Lensed Gravitational Wave Signals}\label{sec:method}
Strong lensing identification methodologies as outlined above have created a framework to answer the question of: ``\textit{for a given pair of images, are these images lensed?}'' but have generally not answered the question of ``\textit{by what}?''. In large part this is because the search methodologies are model-agnostic as to not misidentify an event pair because the method assumes an incorrect model. Some initial steps towards answering the question of the lensing object have been made, however, such as in~\cite{janquart_catalog_selection} where the authorial suggestion was to reweight the detection statistic using model information from various catalogs built from differing models such as those found in~\cite{haris_mgal},~\cite{more_mgal}, or~\cite{wierda_rgal}. Similarly, the methodology presented in~\cite{hanabi_paper} describes the inclusion of the model for the calculation of selection effects to achieve a more complete Bayes factor in the context of ``lensed vs unlensed''. Whilst the first method can deal with more realistic models, it requires building an extended catalog when one wants to explore a realistic lens model, assuming particular source and lens populations and is therefore more prone to systematics than the direct application of a lens model. On the other hand, for the second method, it requires an analytical expression for the magnification probability distribution, which reduces the model it can consider. We note that the latter method could also make use of the results coming from a catalog as a model for the magnification probability distribution but it would then suffer the same caveats as the method presented in~\cite{janquart_catalog_selection}. In both cases, the efforts to use these methods on candidate lensed signals is detailed in~\citet{o3_technical}.

An additional complication for more detailed analysis of strong lensing signals is the need for the detection and identification of multiple signals. For instance, analysis of individual images using a specific lens model as outlined in~\cite{gravelamps_paper} in the geometric optics case does not yield successful differentation between lens models due to the complete degeneracies between the effects of strong lensing and other binary source parameters when only a single image is considered. 

A means of performing model selection based on the output of these detection pipelines, including the multiple images and that does not require resampling or the construction of extended catalogs which requires assumptions on population models, would be beneficial to future lensing searches, and it is such a methodology that we present here. 

The goal of this methodology is to find the evidence for a given lens model, here termed $\mathcal{H}_{\mathrm{mod}}$, which consists of a set of lens parameters, $\bm{\Psi}$. By direct application of Bayes' theorem, the evidence for that model is given by:
\begin{multline}
	p\left(d_{1},d_{2}|\mathcal{H}_{\mathrm{mod}}\right) = \\
	\frac{p\left(d_{1},d_{2}|\bm{\Theta}, \bm{\Phi}, \bm{\Psi}, \mathcal{H}_{\mathrm{mod}}\right)
	      p\left(\bm{\Theta}, \bm{\Phi}, \bm{\Phi}|\mathcal{H}_{\mathrm{mod}}\right)}
	      {p\left(\bm{\Theta}, \bm{\Phi}, \bm{\Phi}|d_{1}, d_{2}, \mathcal{H}_{\mathrm{mod}}\right)}.
	\label{eq:bayes-model-evidence}
\end{multline}

To simplify the following, we here define the evidence for a chosen model as $\mathcal{Z} = p\left(d_{1}, d_{2}|\mathcal{H}_{\mathrm{mod}}\right)$ and the model likelihood as $\mathcal{L}\left(\mathcal{H}_{\mathrm{mod}}\right) = p\left(d_{1}, d_{2} | \bm{\Theta}, \bm{\Phi}, \bm{\Psi}, \mathcal{H}_{\mathrm{mod}}\right)$. We consider calculation of the inverse evidence which, as will be made clear, is easier to solve and once solved may be trivially inverted back to the evidence. This is given from the above as

\begin{equation}
	\frac{1}{\mathcal{Z}} =
	\frac{p\left( \bm{\Theta}, \bm{\Phi}, \bm{\Psi} | d_{1}, d_{2}, \mathcal{H}_{\mathrm{mod}} \right)}
	     {\mathcal{L}\left(\mathcal{H}_{\mathrm{mod}}\right) 
	     p\left(\bm{\Theta}, \bm{\Phi}, \bm{\Psi} | \mathcal{H}_{\mathrm{mod}} \right)}.
	\label{eq:inverse-evidence}
\end{equation}

The posterior forming the numerator of Eq.~\eqref{eq:inverse-evidence} may be expanded further using the chain rule to yield
\begin{multline}
	p\left( \bm{\Theta}, \bm{\Phi}, \bm{\Psi} | d_{1}, d_{2}, \mathcal{H}_{\mathrm{mod}} \right) = \\
	p\left( \bm{\Psi} | \bm{\Theta}, \bm{\Phi}, d_{1}, d_{2}, \mathcal{H}_{\mathrm{mod}} \right) \\
	\times p\left( \bm{\Theta}, \bm{\Phi} | d_{1}, d_{2}, \mathcal{H}_{\mathrm{mod}} \right).
	\label{eq:posterior-exapnsion}
\end{multline}

The lattermost term of Eq.~\eqref{eq:posterior-exapnsion} is in fact, insensitive to the lens model since the apparent lensing parameters and the effective BBH parameters may be fully determined from the data. As such, $p\left(\bm{\Theta}, \bm{\Phi} | d_{1}, d_{2}, \mathcal{H}_{\mathrm{mod}}\right) \equiv p\left( \bm{\Theta}, \bm{\Phi} | d_{1}, d_{2} \right)$ and thus, Eq.~\eqref{eq:inverse-evidence} becomes

\begin{equation}
	\frac{1}{\mathcal{Z}} =
	\frac{p\left(\bm{\Psi} | \bm{\Theta}, \bm{\Phi}, d_{1}, d_{2}, \mathcal{H}_{\mathrm{mod}}\right)
	      p\left(\bm{\Theta}, \bm{\Phi} | d_{1}, d_{2} \right)}
	     {\mathcal{L}\left(\mathcal{H}_{\mathrm{mod}}\right)
	     p\left(\bm{\Theta}, \bm{\Phi}, \bm{\Psi} | \mathcal{H}_{\mathrm{mod}}\right)}.
	\label{eq:inverse-evidence-extended}
\end{equation}

This may be solved by sampling $p\left(\bm{\Theta}, \bm{\Phi}|d_{1}, d_{2}\right)$, computing the remaining terms for that sample and averaging the ratio over all samples, i.e.

\begin{equation}
	\frac{1}{\mathcal{Z}} = \left<
	\frac{p\left(\bm{\Psi} | \bm{\Theta}, \bm{\Phi}, d_{1}, d_{2}, \mathcal{H}_{\mathrm{mod}}\right)}
		     {\mathcal{L}\left(\mathcal{H}_{\mathrm{mod}}\right)
		     p\left( \bm{\Theta}, \bm{\Phi}, \bm{\Psi} | \mathcal{H}_{\mathrm{mod}} \right)}
	\right>_{p\left( \bm{\Theta}, \bm{\Phi} | d_{1}, d_{2} \right)}
	\label{eq:inverse-evidence-sample-average}
\end{equation}

The quantity over which we sample, $p\left(\bm{\Theta}, \bm{\Phi} | d_{1}, d_{2}\right)$, is the output of a model-independent joint parameter estimation pipeline (whether using a full joint parameter estimation or a GOLUM style approach). In addition, the prior and likelihood are known and easily calculable, with the joint likelihood being calculated using the implementation from~\cite{golum_update_paper} with the produced lensed waveform. Finally, the numerator---the conditional probability of the lens parameters given the model---may be computed from the relations between the observables parameters and the model parameters as outlined in Sections~\ref{subsubsec:point-mass} and~\ref{subsubsec:sis}.

As a result, Eq.~\eqref{eq:inverse-evidence-sample-average} offers a means by which to compute the evidence for a given lens model directly from the output of the pipelines that would be used to claim a lensing detection. It also bypasses the need for sampling an extended parameter space meaning that the computation would be extremely rapid upon the completion of the model-agnostic search.

We implement Eq.~\eqref{eq:inverse-evidence-sample-average} into the \textsc{Gravelamps} package~\citep{gravelamps_software} for the models outlined above in which $\bm{\Psi}$ reduces to the redshifted lens mass, $M_{Lz}$, and the dimensionless source position, $y$. Whilst these models are relatively simple, they allow the testing of the methodolgy in a controlled environment in which the observable to lens parameter conversion may be analytically calculated and therefore may be more closely monitored when testing the methodology. However, the implementation within \textsc{Gravelamps} will allow for extending the method to more complex realistic models leveraging its already extant capabilities to rapidly compute lens amplification factors~\citep{gravelamps_paper}.

\section{Investigation of Methodological Perforamnce}\label{sec:testing}
In order to investigate the performance of the methodology in real-world scenarios, it has been subjected to a number of tests. The primary source of investigation was the performance of the methodology on a constructed set of 75 lensed GW signal pairs with known parameters. This allowed investigation of the ability of the method to identify the lensing model as well as the recovery of the model-specific lensing parameters. The stability of the evidence calculation was also investigated by looking at the results of varying the number of samples considered. Finally, whilst a real lensing event remains unavailable, examination of the methodology on real-world data was performed by considering the event pair GW191230--LGW200104: the event pair that was determined by the lensing searches in~\citet{o3_technical} to be the highest significance of the ultimately discarded candidates. 

\subsection{Injection Set Investigation}

The main test of methodological performance was to investigate the results of subjecting a series of simulated lensing events to the method. The testing set consisted of 75 pairs of lensed GW signals. To consider a realistic testing set, the mass, spin, and redshift priors for these were chosen to reflect the inferred population from the GWTC-3 data, i.e.~in-line with~\cite{o3_rates_and_pops}\footnote{To be precise, we took the maximum likelihood parameters of the population models given in~\citep{GWTC3_pop_dataRelease}.}. The other parameters were selected from the usual priors on the parameters, see~\cite{golum_paper} for example. The events were chosen to be lensed by SIS lenses with a uniform prior on the redshifted lens mass between $10^{6} M_{\odot}$ and $10^{9} M_{\odot}$ and a power-law prior on the source position with $\alpha = 1$ over the range of source positions resulting in multiple images, avoiding the direct hit and caustic cases, i.e. $y=0$ and $y=1$ respectively. 

Once the sample source and lens parameters were drawn, the lensed pair signals were constructed and injected into a detector network consisting of the two LIGO and the Virgo detectors with noise representative of the expectations for O4. These were then subject to joint parameter estimation using the \textsc{GOLUM} pipeline~\citep{golum_paper, golum_update_paper} operating with the \textsc{Nessai}~\citep{nessai} nested sampler and the following priors on the model-agnostic lensing observables:

\begin{itemize}
	\item{\textit{Relative Magnification}: Uniform between 0.01 and 50}
	\item{\textit{Time Delay}: Uniform over the range of the injected time delay $\pm$ 0.2 seconds}
	\item{\textit{Morse Index}: Uniform on the discrete values of 0, 0.5, and 1}
\end{itemize}
The signal injection and recovery were done using the \textit{IMRPhenomXPHM} waveform~\citep{IMRPhenomXPHM_paper}. The resulting samples were then subjected to the methodology and the evidences for each of the models compared to arrive at the resultant Bayes factor. 

\begin{figure}
	\includegraphics[width=\linewidth]{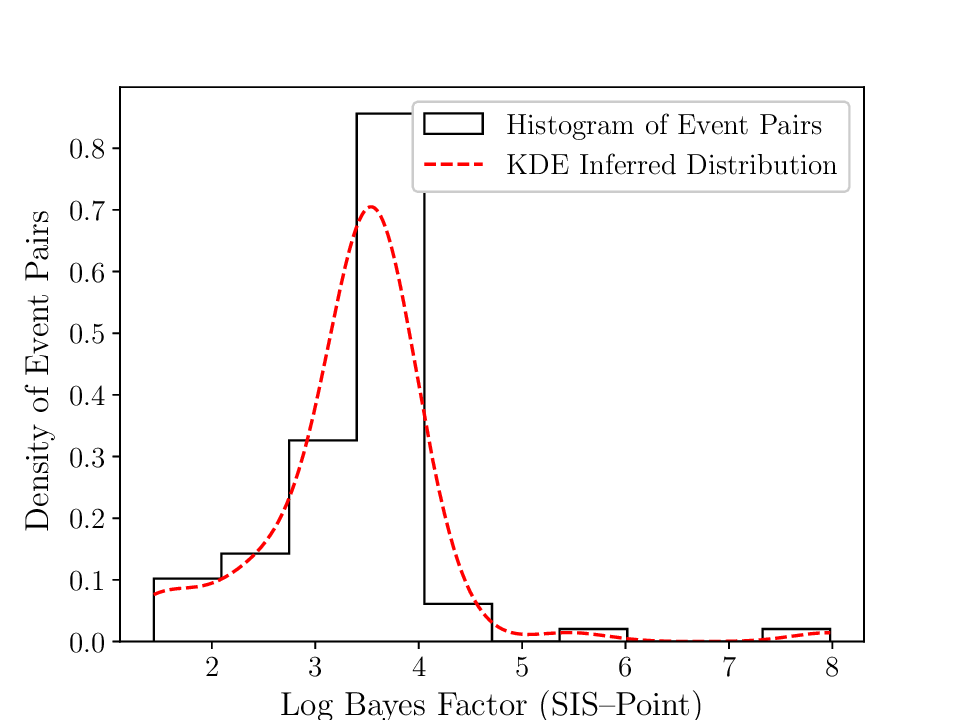}
	\caption{Distribution of the log Bayes factors comparing the SIS and point mass lensing models for the injection set. Shown in black is the raw histogram of these with the red overlay showing the inferred distribution from a kernel-density estimator. All events analysed show a positive Bayes factor which indicates that the SIS was correctly preferred in all cases. }\label{fig:bayes-factor-distribution}
\end{figure}

Figure~\ref{fig:bayes-factor-distribution} shows the distribution of the recovered Bayes factors comparing the SIS and point mass lensing models in the form of both the raw histogram of the data as well as the inferred distribution constructed using a kernel-density estimator (KDE). All of the events considered demonstrate a positive log Bayes factor, indicating a consistent preference for the true SIS model. The minimum log Bayes factor of the considered events was 1.44, the maximum was 7.98, and the average log Bayes factor for the true SIS model across all of the events was 3.37. This is sufficiently large to demonstrate a consistent identification and preference for the correct model in these injections. 

\begin{figure*}
	\centering
	\includegraphics[width=0.49\linewidth]{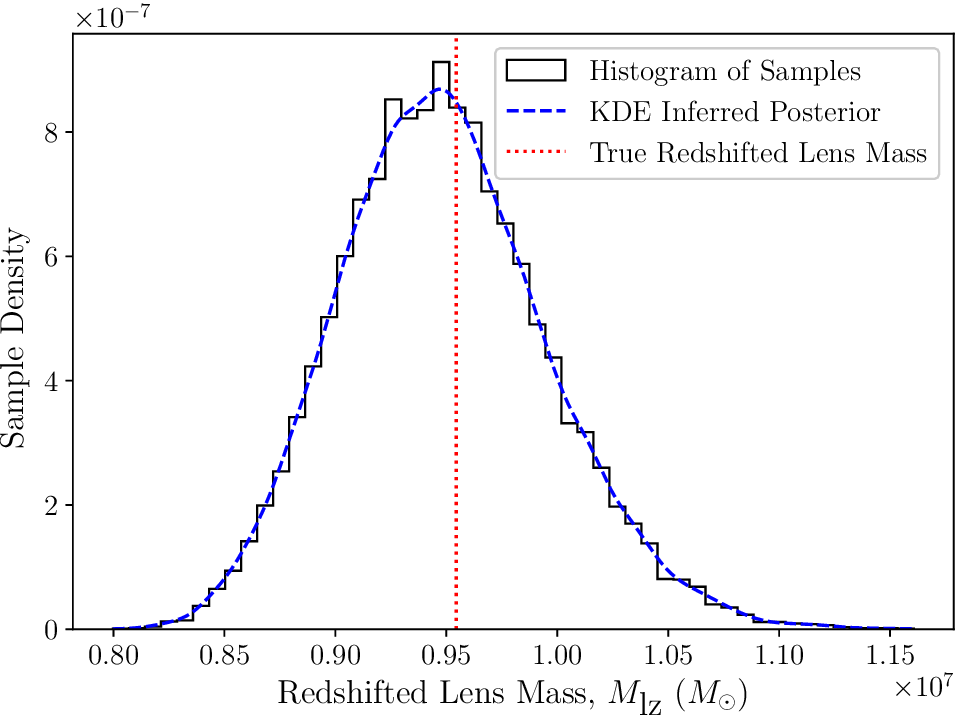}
	\includegraphics[width=0.49\linewidth]{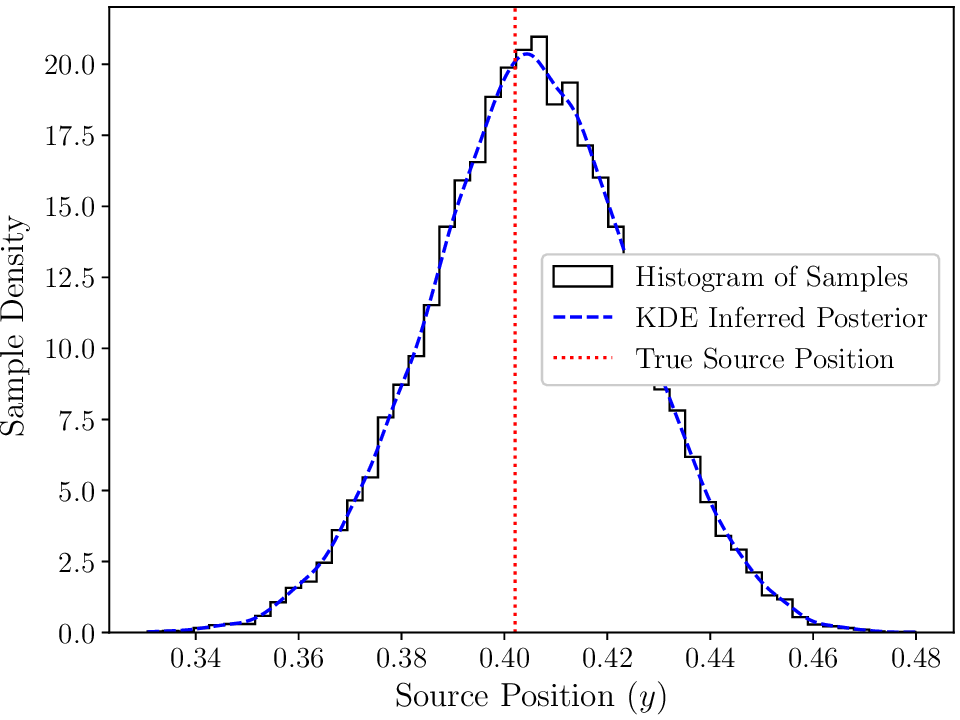}
	\caption{Constructed candidate posteriors on the redshifted lens mass (left) and source position (right). The raw histogram of the samples is presented in black, the inferred posterior distribution using a KDE in blue, and the vertical red dotted line illustrates the true source position value. In this instance, the posteriors indicate a tight constraint with the true values of the parameters inside of the posterior and close to its mode---a successful recovery.}\label{fig:joint-parameter-posteriors}
\end{figure*}

In addition to the model identification, the method produces candidate posteriors for the lens parameters in each of the models. These two may be interrogated to assess the ability of the method to constrain and recover these model specific parameters. Figure~\ref{fig:joint-parameter-posteriors} illustrates an example of the recovered posteriors for the parameters of a single event from the testing set for the SIS model, with the true value indicated. In that case, the parameters demonstrate relatively tight constraints with the true value inside of the constructed posterior. 

\begin{figure*}
	\centering
	\includegraphics[width=\linewidth]{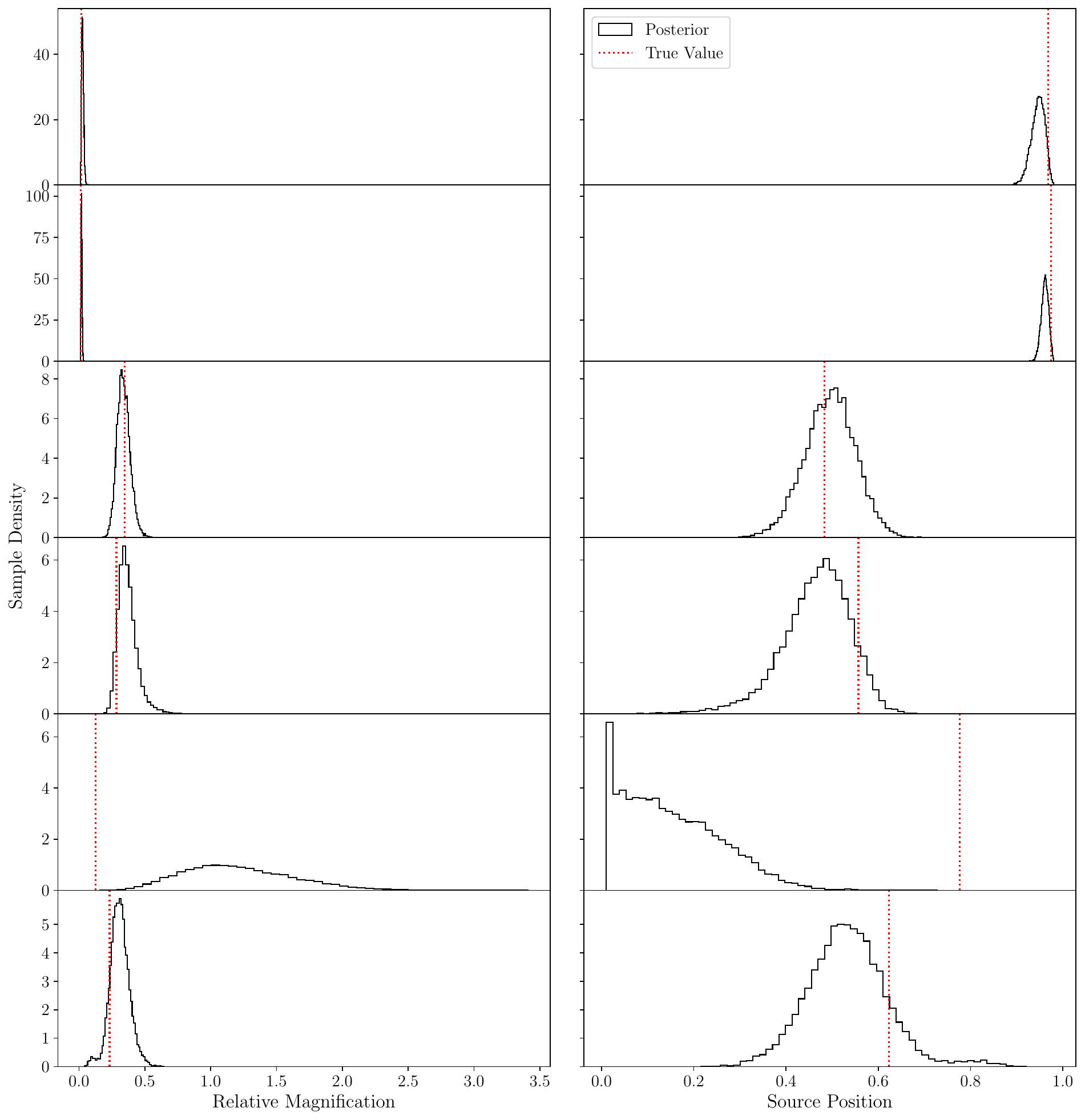}
	\caption{Comparisons between the posterior on the relative magnification as determined by the \textsc{GOLUM} pipeline and the constructed posterior on the source position for the SIS model using the method presented in this work. The true value of these parameters is indicated by the red vertical dotted line. As can be seen, the constraint on the model parameter is dependent upon both the width and accuracy of the lensing observable constraint.}\label{fig:posterior-comparisons}
\end{figure*}

As expected, the tightness and accuracy of the constraint of the model-specific lensing parameters are correlated with the tightness and accuracy of the constraint of the model-agnostic lensing observables. This is demonstrated in Figure~\ref{fig:posterior-comparisons} which contains a subset of the total runs. In this figure, it can be seen that cases with relatively tight constraint of the relative magnification yield relatively tight constrain of the source position. Similarly in the case of the fifth pairing in the figure, the failure to constrain the relative magnification accurately in the model agnostic search results in a failure to constrain the source position accurately. 

\subsection{Stability of Methodological Performance}

Whilst the methodology is relatively inexpensive to perform over all the samples from a given \textsc{GOLUM} investigation, preliminary results may be sought more quickly by using fewer samples. Similarly, the number of samples will differ between investigation of one event and the next, therefore one metric that needs to be investigated for the method is the number of samples that will yield a stable result.

\begin{figure}[t]
	\centering
	\includegraphics[width=\linewidth]{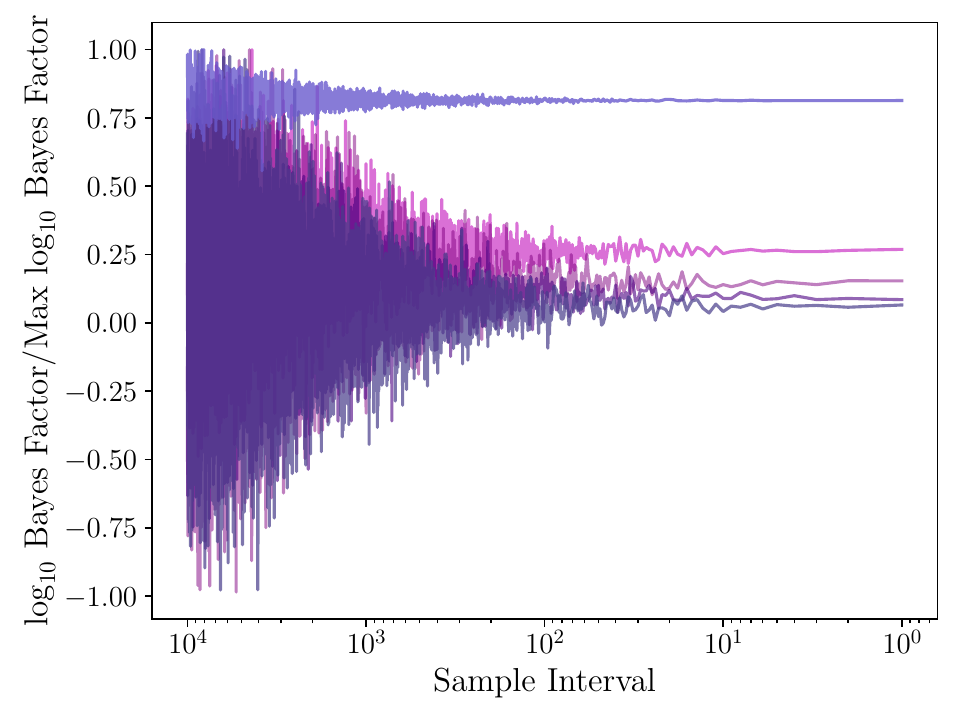}
	\caption{The resulting value of the Bayes factors found by using every $n^{\textrm{th}}$ sample of the full set, scaled by the maximal value from the calculations to allow multiple events to be shown. As can be seen, a preliminary estimate of the Bayes factor, from applying the methodology using every 100$^{\mathrm{th}}$ sample, would be robustly representative of the ``final'' estimate using all of the samples.}\label{fig:stability-test}
\end{figure}

Figure~\ref{fig:stability-test} shows the evolution of the Bayes factor calculation from a number of events as progressively more samples are included. This was done by taking the full sample set and calculating the Bayes factor from every $n^{\textrm{th}}$ sample scaled by the maximal value calculated to allow multiple events to be overlaid. As can be seen from the figure, the estimated value of the Bayes factor shows significant fluctuation when a small number of samples are used and an increasing narrowness as a greater number of samples are included, demonstrating that the solution is stabilising, with an approximate preliminary value being representative at the inclusion of every $100^{\textrm{th}}$ sample.

\begin{figure}[t]
	\centering
	\includegraphics[width=\linewidth]{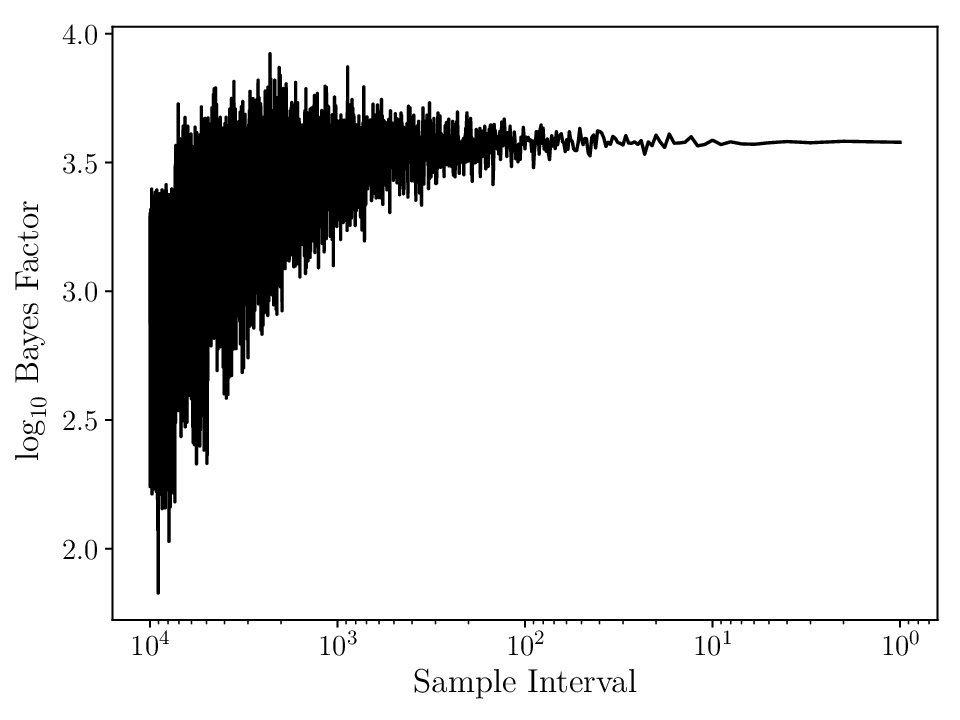}
	\caption{The results of the stability analysis for an event which demonstrates an initial deviation from the later stabilisation value. Similarly to the main set, by approximately the inclusion of every $100^{\textrm{th}}$ sample, the preliminary result would be representative of the final result}\label{fig:stability-deviation}
\end{figure}

In addition to the overall fluctuation of the Bayes factor, a number of events demonstrated that at low sampling rates, the estimated value of the Bayes factor can appear to cluster around a value that is different from the value to which the estimate later appears to converge to. An example of this is shown in Figure~\ref{fig:stability-deviation}. Similarly, the figure shows an approximate point at which a preliminary result---from the inclusion of a limited number of samples---is robustly representative of the true Bayes factor. Our results again would indicate that including every 100$^{\mathrm{th}}$ sample is sufficiently representative of the Bayes factor in this way. 

In addition to the overall fluctuation of the Bayes factor, a number of events demonstrated that at low sampling rates, the estimated value of the Bayes factor would appear to cluster around a deviated value from the later settling value. An example of this is shown in Figure~\ref{fig:stability-deviation}. Similarly, this would place an approximate point at which a prelimninary result would be representative of the final value at the inclusion of every $100^{\textrm{th}}$ sample. 

\subsection{Example Deployment on GW191230--LGW200104}

As discussed in the introduction, as of the end of the third observing run, the LVK lensing searches which include joint parameter estimation analyses that would be appropriate for this method have yet to yield any confirmed detections~\citep{o2_lensing, o3a_lensing, o3_lensing}. Consequently, there is no real lensing data on which to perform a test case analysis. To represent such a test case, we deploy the method on the event pair GW191230--LGW200104 which was identified to have the highest Bayes factor of the ultimately discarded candidates from searches thus far~\citep{o3_lensing, o3_technical}. Similarly to~\cite{o3_technical}, we stress that we do not claim that this pair is a genuine lensing event. However, we will treat it as though it were for the purposes of this test deployment. Similarly, due to the relatively high SNR of the trigger, LGW200104 it is treated as a real GW event despite the very low $p_{\textrm{astro}}$ of $1\%$ which would suggest this is unlikely in actuality~\citep{o3_technical}.

\begin{figure*}[t]
	\centering
	\includegraphics[width=\linewidth]{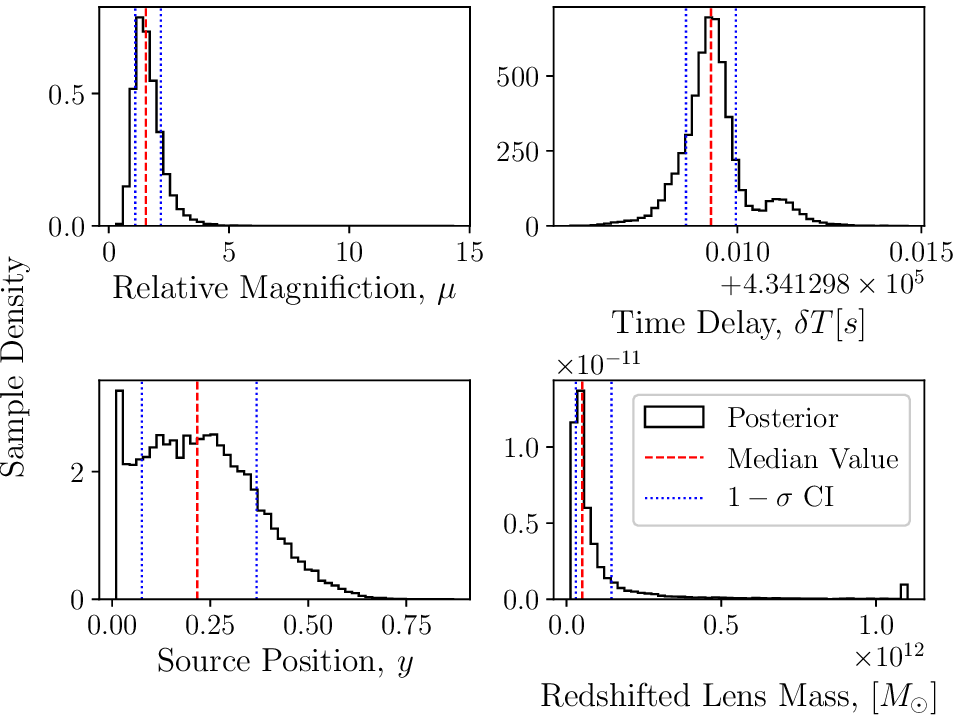}
	\caption{\textit{Top}: Posteriors on the relative lensing parameters for the GW191230--LGW200104 pair as determined by the GOLUM investigation.\ \textit{Bottom}: The reconstructed posteriors for the source position and redshifted lens mass as calculated by the methodology presented in this work.}\label{fig:super-sub-posteriors}
\end{figure*}

The pair has been re-analysed using the GOLUM pipeline and the \textsc{Nessai} nested sampler in order to yield an equivalent model-agnostic search result to those of the injection set. Performing the model selection method on the results yielded a result that favours the SIS model with a $\log_{10}$ Bayes factor of 3.65. This is consistent with the finding of~\cite{o3_technical} that the consideration of the SIS model did improve preference for the pair as compared to the raw model-agnostic investigation---though~\cite{o3_technical} had the greatest improvement with the SIE model. The posteriors from the GOLUM investigation as well as the reconstructed posteriors for the SIS parameters are shown in Figure~\ref{fig:super-sub-posteriors}. These are fairly broad as might be expected when including a trigger that is likely not a genuine lensing event or indeed a genuine GW event. Additionally, we note that there is distinct railing in the reconstructed source position posterior against the direct hit---i.e.\ $y=0$---case which may too be an indication of the probably unlensed nature of the candidate pair. However, should this have been a genuine lensed event pair the median values of these reconstructed parameter posteriors for the SIS model would suggest a $5.08^{+0.94}_{-2.09} \times 10^{10} M_{\odot}$ lens at a source position of $0.23^{+0.15}_{-0.14}$.

\section{Conclusion}\label{sec:conclusion}
With the increasing sensitivity of the interferometers, the detection of gravitational lensing of a GW event becomes increasingly likely. When such a detection occurs, the nature of the lens, as well as the source, will be studied. One of the fundamental questions about the lens will be about how the mass is distributed. We present here a method by which to rapidly determine the mass density profile using the output of the model-agnostic strong lensing joint parameter estimation pipelines without the need to build extended catalogs of lenses, under the assumption of source and lens populations. This method has been implemented within the \textsc{python} package \textsc{Gravelamps}~\citep{gravelamps_software} to augment its pre-existing capabilities to determine mass density profiles for microlensing events.

We have demonstrated the efficacy of the method using a pair of relatively simple lens models---the point mass and the SIS lenses---though it is not restricted to these models. In order to test the methodology's performance it was confronted with an injected set of 75 lensed BBH pairs following the observed distribution of events during the third observing run of the LVK detectors. In each case, the correct lensing model was identified with the lens model-specific parameter recovery falling in line with the recovery of the observable model-agnostic lensing parameters.

To test the stability of the results calculated, we observed the behaviour of the Bayes factors as more and more samples are included. Preliminary results from the methodology would be representative of the final solution using all of the samples from the inclusion of every $100^{\textrm{th}}$ sample.

Finally, whilst there remains no real lensing data on which to deploy the methodology, we test deployment on actual data by examining the GW191230--LGW200104 pair. It had the highest Bayes factor of the ultimately discarded pairs. Our analysis indicated that if the pair were genuinely lensed, which we stress is unlikely, then the preferred lens of the two models examined would be a $5.08^{+0.94}_{-2.09} \times 10^{10}M_{\odot}$ SIS lens at a dimensionless displacement from the optical axis of $0.23^{+0.15}_{-0.14}$. This is in line with the increased support for the pair when considering the SIS model in~\cite{o3_technical}---though we note that the SIE model was the most preferred in that case. 

With the demonstration of viability for the method, future work to be undertaken would be to investigate more realistic models for the mass density profile of the lens, such as the aforementioned SIE, or the Navarro-Frenk-White model. For these more sophisticated models, additional work would need to be done to rapidly go from the lensing observables to the model-specific parameters without the relatively simple analytical relationships that exist for the point mass and SIS models considered in this work.

\begin{acknowledgements}
\section*{Acknowledgements}\label{sec:acknowledgements}
The authors thank Chris Van Den Broeck for useful discussions and insights on related topics. MW acknowledges the support of the Science and Technologies Facilities Council (STFC) of the United Kingdom. JJ is supported by the research programme of the Netherlands Organisation for Scientific Research (NWO). MH acknowledges additional support from the Science and Technologies Facilities Council (Ref. ST/L000946/1). 

This material is based upon work supported by NSF's LIGO Laboratory which is a major facility fully funded by the National Science Foundation.

The authors are grateful for computational resources provided by the LIGO Laboratory and the Leonard E Parker Center for Gravitation, Cosmology, and Astrophysics at the University of Wisconsin-Milwaukee and supported by National Science Foundation Grants PHY-0757058 and PHY-0823459, and PHY-1626190 and PHY-1700765 respectively.

\end{acknowledgements}

\pagebreak

\bibliographystyle{aasjournal}
\bibliography{grave_golum}

\begin{thebibliography}{}
\expandafter\ifx\csname natexlab\endcsname\relax\def\natexlab#1{#1}\fi
\providecommand{\url}[1]{\href{#1}{#1}}
\providecommand{\dodoi}[1]{doi:~\href{http://doi.org/#1}{\nolinkurl{#1}}}
\providecommand{\doeprint}[1]{\href{http://ascl.net/#1}{\nolinkurl{http://ascl.net/#1}}}
\providecommand{\doarXiv}[1]{\href{https://arxiv.org/abs/#1}{\nolinkurl{https://arxiv.org/abs/#1}}}

\bibitem[{Aasi {et~al.}(2015)}]{advancedligo}
Aasi, J., {et~al.} 2015, Class. Quant. Grav., 32, 074001, \dodoi{10.1088/0264-9381/32/7/074001}

\bibitem[{Abbott {et~al.}(2021)Abbott, Abbott, Abraham, Acernese, Ackley, Adams, Adams, Adhikari, Adya, Affeldt, Agarwal, Agathos, Agatsuma, Aggarwal, Aguiar, Aiello, Ain, Ajith, Aleman, Allen, Allocca, Altin, Amato, Anand, Ananyeva, Anderson, Anderson, Angelova, Ansoldi, Antelis, Antier, Appert, Arai, Araya, Areeda, Arène, Arnaud, Aronson, Arun, Asali, Ashton, Aston, Astone, Aubin, Aufmuth, AultONeal, Austin, Babak, Badaracco, Bader, Bae, Baer, Bagnasco, Bai, Baird, Ball, Ballardin, Ballmer, Bals, Balsamo, Baltus, Banagiri, Bankar, Bankar, Barayoga, Barbieri, Barish, Barker, Barneo, Barone, Barr, Barsotti, Barsuglia, Barta, Bartlett, Barton, Bartos, Bassiri, Basti, Bawaj, Bayley, Baylor, Bazzan, Bécsy, Bedakihale, Bejger, Belahcene, Benedetto, Beniwal, Benjamin, Bennett, Bentley, BenYaala, Bergamin, Berger, Bernuzzi, Berry, Bersanetti, Bertolini, Betzwieser, Bhandare, Bhandari, Bhattacharjee, Bhaumik, Bidler, Bilenko, Billingsley, Birney, Birnholtz, Biscans, Bischi, Biscoveanu, Bisht, Biswas, Bitossi, Bizouard, Blackburn, Blackman, Blair, Blair, Blair, Bobba, Bode, Boer, Bogaert, Boldrini, Bondu, Bonilla, Bonnand, Booker, Boom, Bork, Boschi, Bose, Bose, Bossilkov, Boudart, Bouffanais, Bozzi, Bradaschia, Brady, Bramley, Branch, Branchesi, Brau, Breschi, Briant, Briggs, Brillet, Brinkmann, Brockill, Brooks, Brooks, Brown, Brunett, Bruno, Bruntz, Bryant, Buikema, Bulik, Bulten, Buonanno, Buscicchio, Buskulic, Byer, Cadonati, Caesar, Cagnoli, Cahillane, III, Bustillo, Callaghan, Callister, Calloni, Camp, Canepa, Cannavacciuolo, Cannon, Cao, Cao, Capote, Carapella, Carbognani, Carlin, Carney, Carpinelli, Carullo, Carver, Diaz, Casentini, Castaldi, Caudill, Cavaglià, Cavalier, Cavalieri, Cella, Cerdá-Durán, Cesarini, Chaibi, Chakravarti, Champion, Chan, Chan, Chan, Chandra, Chanial, Chao, Charlton, Chase, Chassande-Mottin, Chatterjee, Chaturvedi, Chen, Chen, Chen, Chen, Chen, Chen, Cheng, Cheong, Cheung, Chia, Chiadini, Chierici, Chincarini, Chiofalo, Chiummo, Cho,
  Cho, Choate, Choudhary, Choudhary, Christensen, Chu, Chua, Chung, Ciani, Ciecielag, Cieślar, Cifaldi, Ciobanu, Ciolfi, Cipriano, Cirone, Clara, Clark, Clark, Clarke, Clearwater, Clesse, Cleva, Coccia, Cohadon, Cohen, Cohen, Colleoni, Collette, Colpi, Compton, Constancio, Conti, Cooper, Corban, Corbitt, Cordero-Carrión, Corezzi, Corley, Cornish, Corre, Corsi, Cortese, Costa, Cotesta, Coughlin, Coughlin, Coulon, Countryman, Cousins, Couvares, Covas, Coward, Cowart, Coyne, Coyne, Creighton, Creighton, Criswell, Croquette, Crowder, Cudell, Cullen, Cumming, Cummings, Cuoco, Curyło, Canton, Dálya, Dana, DaneshgaranBajastani, D’Angelo, Danilishin, D’Antonio, Danzmann, Darsow-Fromm, Dasgupta, Datrier, Dattilo, Dave, Davier, Davies, Davis, Daw, Dean, DeBra, Deenadayalan, Degallaix, Laurentis, Deléglise, Favero, Lillo, Lillo, Pozzo, DeMarchi, Matteis, D’Emilio, Demos, Dent, Depasse, Pietri, Rosa, Rossi, DeSalvo, Simone, Dhurandhar, Díaz, Diaz-Ortiz, Didio, Dietrich, Fiore, Fronzo, Giorgio, Giovanni, Girolamo, Lieto, Ding, Pace, Palma, Renzo, Divakarla, Dmitriev, Doctor, D’Onofrio, Donovan, Dooley, Doravari, Dorrington, Drago, Driggers, Drori, Du, Ducoin, Dupej, Durante, D’Urso, Duverne, Dwyer, Easter, Ebersold, Eddolls, Edelman, Edo, Edy, Effler, Eichholz, Eikenberry, Eisenmann, Eisenstein, Ejlli, Errico, Essick, Estellés, Estevez, Etienne, Etzel, Evans, Evans, Ewing, Ezquiaga, Fafone, Fair, Fairhurst, Fan, Farah, Farinon, Farr, Farr, Farrow, Fauchon-Jones, Favata, Fays, Fazio, Feicht, Fejer, Feng, Fenyvesi, Ferguson, Fernandez-Galiana, Ferrante, Ferreira, Fidecaro, Figura, Fiori, Fishbach, Fisher, Fittipaldi, Fiumara, Flaminio, Floden, Flynn, Fong, Font, Fornal, Forsyth, Franke, Frasca, Frasconi, Frederick, Frei, Freise, Frey, Fritschel, Frolov, Fronzé, Fulda, Fyffe, Gabbard, Gadre, Gaebel, Gair, Gais, Galaudage, Gamba, Ganapathy, Ganguly, Gaonkar, Garaventa, García-Núñez, García-Quirós, Garufi, Gateley, Gaudio, Gayathri, Gemme, Gennai, George,
  Gergely, Gewecke, Ghonge, Ghosh, Ghosh, Ghosh, Ghosh, Ghosh, Giacomazzo, Giacoppo, Giaime, Giardina, Gibson, Gier, Giesler, Giri, Gissi, Glanzer, Gleckl, Godwin, Goetz, Goetz, Gohlke, Goncharov, González, Gopakumar, Gosselin, Gouaty, Goyal, Grace, Grado, Granata, Granata, Grant, Gras, Grassia, Gray, Gray, Greco, Green, Green, Gretarsson, Gretarsson, Griffith, Griffiths, Griggs, Grignani, Grimaldi, Grimes, Grimm, Grote, Grunewald, Gruning, Guerrero, Guidi, Guimaraes, Guixé, Gulati, Guo, Guo, Gupta, Gupta, Gupta, Gustafson, Gustafson, Guzman, Haegel, Halim, Hall, Hamilton, Hammond, Haney, Hanks, Hanna, Hannam, Hannuksela, Hansen, Hansen, Hanson, Harder, Hardwick, Haris, Harms, Harry, Harry, Hartwig, Haskell, Hasskew, Haster, Haughian, Hayes, Healy, Heidmann, Heintze, Heinze, Heinzel, Heitmann, Hellman, Hello, Helmling-Cornell, Hemming, Hendry, Heng, Hennes, Hennig, Hennig, Vivanco, Heurs, Hild, Hill, Hines, Hochheim, Hofman, Hohmann, Holgado, Holland, Hollows, Holmes, Holt, Holz, Hopkins, Hough, Howell, Hoy, Hoyland, Hreibi, Hsu, Huang, Hübner, Huddart, Huerta, Hughey, Hui, Husa, Huttner, Huxford, Huynh-Dinh, Idzkowski, Iess, Inchauspe, Ingram, Intini, Isi, Isleif, Iyer, JaberianHamedan, Jacqmin, Jadhav, Jadhav, James, Jan, Jani, Janquart, Janssens, Janthalur, Jaranowski, Jariwala, Jaume, Jenkins, Jeunon, Jia, Jiang, Johns, Jones, Jones, Jones, Jones, Jones, Jonker, Ju, Junker, Kalaghatgi, Kalogera, Kamai, Kandhasamy, Kang, Kanner, Kao, Kapadia, Kapasi, Karat, Karathanasis, Karki, Kashyap, Kasprzack, Kastaun, Katsanevas, Katsavounidis, Katzman, Kaur, Kawabe, Kéfélian, Keitel, Key, Khadka, Khalili, Khan, Khan, Khazanov, Khetan, Khursheed, Kijbunchoo, Kim, Kim, Kim, Kim, Kim, Kimball, King, Kinley-Hanlon, Kirchhoff, Kissel, Kleybolte, Klimenko, Knee, Knowles, Knyazev, Koch, Koekoek, Koley, Kolitsidou, Kolstein, Komori, Kondrashov, Kontos, Koper, Korobko, Kovalam, Kozak, Kringel, Krishnendu, Królak, Kuehn, Kuei, Kumar, Kumar, Kumar, Kumar, Kuns, Kwang, Laghi,
  Lalande, Lam, Lamberts, Landry, Lane, Lang, Lange, Lantz, Rosa, Lartaux-Vollard, Lasky, Laxen, Lazzarini, Lazzaro, Leaci, Leavey, Lecoeuche, Lee, Lee, Lee, Lee, Lehmann, Lemaître, Leon, Leroy, Letendre, Levin, Leviton, Li, Li, Li, Li, Li, Linde, Linker, Linley, Littenberg, Liu, Liu, Liu, Llorens-Monteagudo, Lo, Lockwood, Lollie, London, Longo, Lopez, Lorenzini, Loriette, Lormand, Losurdo, Lough, Lousto, Lovelace, Lück, Lumaca, Lundgren, Macas, MacInnis, Macleod, MacMillan, Macquet, Hernandez, Magaña-Sandoval, Magazzù, Magee, Maggiore, Majorana, Makarem, Maksimovic, Maliakal, Malik, Man, Mandic, Mangano, Mango, Mansell, Manske, Mantovani, Mapelli, Marchesoni, Marion, Mark, Márka, Márka, Markakis, Markosyan, Markowitz, Maros, Marquina, Marsat, Martelli, Martin, Martin, Martinez, Martinez, Martinovic, Martynov, Marx, Masalehdan, Mason, Massera, Masserot, Massinger, Masso-Reid, Mastrogiovanni, Matas, Mateu-Lucena, Matichard, Matiushechkina, Mavalvala, McCann, McCarthy, McClelland, McClincy, McCormick, McCuller, McGhee, McGuire, McIsaac, McIver, McManus, McRae, McWilliams, Meacher, Mehmet, Mehta, Melatos, Melchor, Mendell, Menendez-Vazquez, Menoni, Mercer, Mereni, Merfeld, Merilh, Merritt, Merzougui, Meshkov, Messenger, Messick, Meyers, Meylahn, Mhaske, Miani, Miao, Michaloliakos, Michel, Middleton, Milano, Miller, Millhouse, Mills, Milotti, Milovich-Goff, Minazzoli, Minenkov, Mir, Mishkin, Mishra, Mishra, Mistry, Mitra, Mitrofanov, Mitselmakher, Mittleman, Mo, Mogushi, Mohapatra, Mohite, Molina, Molina-Ruiz, Mondin, Montani, Moore, Moraru, Morawski, More, Moreno, Moreno, Morisaki, Mours, Mow-Lowry, Mozzon, Muciaccia, Mukherjee, Mukherjee, Mukherjee, Mukherjee, Mukund, Mullavey, Munch, Muñiz, Murray, Musenich, Nadji, Nagar, Nardecchia, Naticchioni, Nayak, Nayak, Neil, Neilson, Nelemans, Nelson, Nery, Neunzert, Ng, Ng, Nguyen, Nguyen, Nguyen, Nichols, Nissanke, Nocera, Noh, Norman, North, Nuttall, Oberling, O’Brien, O’Dell, Oganesyan, Oh, Oh, Ohme, Ohta,
  Okada, Olivetto, Oram, O’Reilly, Ormiston, Ormsby, Ortega, O’Shaughnessy, O’Shea, Ossokine, Osthelder, Ottaway, Overmier, Pace, Pagano, Page, Pagliaroli, Pai, Pai, Palamos, Palashov, Palomba, Panda, Pang, Pankow, Pannarale, Pant, Paoletti, Paoli, Paolone, Parker, Pascucci, Pasqualetti, Passaquieti, Passuello, Patel, Patricelli, Payne, Pechsiri, Pedraza, Pegoraro, Pele, Penn, Perego, Pereira, Pereira, Perez, Périgois, Perreca, Perriès, Petermann, Petterson, Pfeiffer, Pham, Phukon, Piccinni, Pichot, Piendibene, Piergiovanni, Pierini, Pierro, Pillant, Pilo, Pinard, Pinto, Piotrzkowski, Piotrzkowski, Pirello, Pitkin, Placidi, Plastino, Pluchar, Poggiani, Polini, Pong, Ponrathnam, Popolizio, Porter, Powell, Pracchia, Pradier, Prajapati, Prasai, Prasanna, Pratten, Prestegard, Principe, Prodi, Prokhorov, Prosposito, Prudenzi, Puecher, Punturo, Puosi, Puppo, Pürrer, Qi, Quetschke, Quinonez, Quitzow-James, Raab, Raaijmakers, Radkins, Radulesco, Raffai, Rail, Raja, Rajan, Ramirez, Ramirez, Ramos-Buades, Rana, Rapagnani, Rapol, Ratto, Raymond, Raza, Razzano, Read, Rees, Regimbau, Rei, Reid, Reitze, Relton, Rettegno, Ricci, Richardson, Richardson, Richardson, Ricker, Riemenschneider, Riles, Rizzo, Robertson, Robie, Robinet, Rocchi, Rocha, Rodriguez, Rodriguez-Soto, Rolland, Rollins, Roma, Romanelli, Romano, Romel, Romero, Romero-Shaw, Romie, Rose, Rosińska, Rosofsky, Ross, Rowan, Rowlinson, Roy, Roy, Rozza, Ruggi, Ryan, Sachdev, Sadecki, Sadiq, Sakellariadou, Salafia, Salconi, Saleem, Salemi, Samajdar, Sanchez, Sanchez, Sanchez, Sanchis-Gual, Sanders, Sanuy, Saravanan, Sarin, Sassolas, Satari, Sathyaprakash, Sauter, Savage, Savant, Sawant, Sawant, Sayah, Schaetzl, Scheel, Scheuer, Schindler-Tyka, Schmidt, Schnabel, Schneewind, Schofield, Schönbeck, Schulte, Schutz, Schwartz, Scott, Scott, Seglar-Arroyo, Seidel, Sellers, Sengupta, Sennett, Sentenac, Seo, Sequino, Sergeev, Setyawati, Shaffer, Shahriar, Shams, Sharifi, Sharma, Sharma, Shawhan, Shcheblanov, Shen,
  Shikauchi, Shink, Shoemaker, Shoemaker, Shukla, ShyamSundar, Sieniawska, Sigg, Singer, Singh, Singh, Singha, Sintes, Sipala, Skliris, Slagmolen, Slaven-Blair, Smetana, Smith, Smith, Somala, Son, Soni, Soni, Sorazu, Sordini, Sorrentino, Sorrentino, Soulard, Souradeep, Sowell, Spagnuolo, Spencer, Spera, Srivastava, Srivastava, Staats, Stachie, Steer, Steinlechner, Steinlechner, Stops, Stover, Strain, Strang, Stratta, Strunk, Sturani, Stuver, Südbeck, Sudhagar, Sudhir, Suh, Summerscales, Sun, Sun, Sunil, Sur, Suresh, Sutton, Swinkels, Szczepańczyk, Szewczyk, Tacca, Tait, Talbot, Tanasijczuk, Tanner, Tao, Tapia, Martin, Tasson, Tenorio, Terkowski, Test, Thirugnanasambandam, Thomas, Thomas, Thompson, Thondapu, Thorne, Thrane, Tiwari, Tiwari, Tiwari, Toland, Tolley, Tonelli, Torres-Forné, Torrie, e~Melo, Töyrä, Trapananti, Travasso, Traylor, Tringali, Tripathee, Troiano, Trovato, Trudeau, Tsai, Tsai, Tsang, Tse, Tso, Tsukada, Tsuna, Tsutsui, Turconi, Ubhi, Udall, Ueno, Ugolini, Unnikrishnan, Urban, Usman, Utina, Vahlbruch, Vajente, Vajpeyi, Valdes, Valentini, Valsan, van Bakel, van Beuzekom, van~den Brand, Broeck, Vander-Hyde, van~der Schaaf, van Heijningen, Vanosky, Vardaro, Vargas, Varma, Vasúth, Vecchio, Vedovato, Veitch, Veitch, Venkateswara, Venneberg, Venugopalan, Verkindt, Verma, Veske, Vetrano, Viceré, Viets, Villa-Ortega, Vinet, Vitale, Vo, Vocca, von Reis, von Wrangel, Vorvick, Vyatchanin, Wade, Wade, Wagner, Walet, Walker, Wallace, Wallace, Walsh, Wang, Wang, Ward, Warner, Was, Washington, Watchi, Weaver, Wei, Weinert, Weinstein, Weiss, Weller, Wellmann, Wen, Weßels, Westhouse, Wette, Whelan, White, Whiting, Whittle, Wilken, Williams, Williams, Williamson, Willis, Willke, Wilson, Winkler, Wipf, Wlodarczyk, Woan, Woehler, Wofford, Wong, Wright, Wu, Wysocki, Xiao, Yamamoto, Yang, Yang, Yang, Yang, Yap, Yeeles, Yelikar, Yeung, Ying, Yoon, Yu, Yu, Zadrożny, Zanolin, Zelenova, Zendri, Zevin, Zhang, Zhang, Zhang, Zhang, Zhao, Zhao, Zhao, Zhou, Zhu,
  Zimmerman, Zucker, Zweizig, Collaboration, \& the Virgo~Collaboration}]{o3a_lensing}
Abbott, R., Abbott, T.~D., Abraham, S., {et~al.} 2021, The Astrophysical Journal, 923, 14, \dodoi{10.3847/1538-4357/ac23db}

\bibitem[{Abbott {et~al.}(2022)Abbott, Abbott, Acernese, Ackley, Adams, Adhikari, Adhikari, Adya, Affeldt, Agarwal, Agathos, Agatsuma, Aggarwal, Aguiar, Aiello, Ain, Ajith, Akutsu, Albanesi, Allocca, Altin, Amato, Anand, Anand, Ananyeva, Anderson, Anderson, Ando, Andrade, Andres, Andrić, Angelova, Ansoldi, Antelis, Antier, Antonini, Appert, Arai, Arai, Arai, Araki, Araya, Araya, Areeda, Arène, Aritomi, Arnaud, Aronson, Arun, Asada, Asali, Ashton, Aso, Assiduo, Aston, Astone, Aubin, Austin, Babak, Badaracco, Bader, Badger, Bae, Bae, Baer, Bagnasco, Bai, Baiotti, Baird, Bajpai, Ball, Ballardin, Ballmer, Balsamo, Baltus, Banagiri, Bankar, Barayoga, Barbieri, Barish, Barker, Barneo, Barone, Barr, Barsotti, Barsuglia, Barta, Bartlett, Barton, Bartos, Bassiri, Basti, Bawaj, Bayley, Baylor, Bazzan, Bécsy, Bedakihale, Bejger, Belahcene, Benedetto, Beniwal, Bennett, Bentley, BenYaala, Bergamin, Berger, Bernuzzi, Berry, Bersanetti, Bertolini, Betzwieser, Beveridge, Bhandare, Bhardwaj, Bhattacharjee, Bhaumik, Bilenko, Billingsley, Bini, Birney, Birnholtz, Biscans, Bischi, Biscoveanu, Bisht, Biswas, Bitossi, Bizouard, Blackburn, Blair, Blair, Blair, Bobba, Bode, Boer, Bogaert, Boldrini, Bonavena, Bondu, Bonilla, Bonnand, Booker, Boom, Bork, Boschi, Bose, Bose, Bossilkov, Boudart, Bouffanais, Bozzi, Bradaschia, Brady, Bramley, Branch, Branchesi, Brau, Breschi, Briant, Briggs, Brillet, Brinkmann, Brockill, Brooks, Brooks, Brown, Brunett, Bruno, Bruntz, Bryant, Bulik, Bulten, Buonanno, Buscicchio, Buskulic, Buy, Byer, Cadonati, Cagnoli, Cahillane, Bustillo, Callaghan, Callister, Calloni, Cameron, Camp, Canepa, Canevarolo, Cannavacciuolo, Cannon, Cao, Cao, Capocasa, Capote, Carapella, Carbognani, Carlin, Carney, Carpinelli, Carrillo, Carullo, Carver, Diaz, Casentini, Castaldi, Caudill, Cavaglià, Cavalier, Cavalieri, Ceasar, Cella, Cerdá-Durán, Cesarini, Chaibi, Chakravarti, Subrahmanya, Champion, Chan, Chan, Chan, Chan, Chan, Chandra, Chanial, Chao, Charlton, Chase,
  Chassande-Mottin, Chatterjee, Chatterjee, Chatterjee, Chaturvedi, Chaty, Chatziioannou, Chen, Chen, Chen, Chen, Chen, Chen, Chen, Chen, Cheng, Cheong, Cheung, Chia, Chiadini, Chiang, Chiarini, Chierici, Chincarini, Chiofalo, Chiummo, Cho, Cho, Choudhary, Choudhary, Christensen, Chu, Chu, Chu, Chua, Chung, Ciani, Ciecielag, Cieślar, Cifaldi, Ciobanu, Ciolfi, Cipriano, Cirone, Clara, Clark, Clark, Clarke, Clearwater, Clesse, Cleva, Coccia, Codazzo, Cohadon, Cohen, Cohen, Colleoni, Collette, Colombo, Colpi, Compton, au2, Conti, Cooper, Corban, Corbitt, Cordero-Carrión, Corezzi, Corley, Cornish, Corre, Corsi, Cortese, Costa, Cotesta, Coughlin, Coulon, Countryman, Cousins, Couvares, Coward, Cowart, Coyne, Coyne, Creighton, Creighton, Criswell, Croquette, Crowder, Cudell, Cullen, Cumming, Cummings, Cunningham, Cuoco, Curyło, Dabadie, Canton, Dall'Osso, Dálya, Dana, DaneshgaranBajastani, D'Angelo, Danilishin, D'Antonio, Danzmann, Darsow-Fromm, Dasgupta, Datrier, Datta, Dattilo, Dave, Davier, Davies, Davis, Davis, Daw, Dean, DeBra, Deenadayalan, Degallaix, Laurentis, Deléglise, Favero, Lillo, Lillo, Pozzo, DeMarchi, Matteis, D'Emilio, Demos, Dent, Depasse, Pietri, Rosa, Rossi, DeSalvo, Simone, Dhurandhar, Díaz, au2, Didio, Dietrich, Fiore, Fronzo, Giorgio, Giovanni, Giovanni, Girolamo, Lieto, Ding, Pace, Palma, Renzo, Divakarla, Dmitriev, Doctor, D'Onofrio, Donovan, Dooley, Doravari, Dorrington, Drago, Driggers, Drori, Ducoin, Dupej, Durante, D'Urso, Duverne, Dwyer, Eassa, Easter, Ebersold, Eckhardt, Eddolls, Edelman, Edo, Edy, Effler, Eguchi, Eichholz, Eikenberry, Eisenmann, Eisenstein, Ejlli, Engelby, Enomoto, Errico, Essick, Estellés, Estevez, Etienne, Etzel, Evans, Evans, Ewing, Fafone, Fair, Fairhurst, Farah, Farinon, Farr, Farr, Farrow, Fauchon-Jones, Favaro, Favata, Fays, Fazio, Feicht, Fejer, Fenyvesi, Ferguson, Fernandez-Galiana, Ferrante, Ferreira, Fidecaro, Figura, Fiori, Fishbach, Fisher, Fittipaldi, Fiumara, Flaminio, Floden, Fong, Font, Fornal,
  Forsyth, Franke, Frasca, Frasconi, Frederick, Freed, Frei, Freise, Frey, Fritschel, Frolov, Fronzé, Fujii, Fujikawa, Fukunaga, Fukushima, Fulda, Fyffe, Gabbard, Gadre, Gair, Gais, Galaudage, Gamba, Ganapathy, Ganguly, Gao, Gaonkar, Garaventa, García-Núñez, García-Quirós, Garufi, Gateley, Gaudio, Gayathri, Ge, Gemme, Gennai, George, Gerberding, Gergely, Gewecke, Ghonge, Ghosh, Ghosh, Ghosh, Ghosh, Giacomazzo, Giacoppo, Giaime, Giardina, Gibson, Gier, Giesler, Giri, Gissi, Glanzer, Gleckl, Godwin, Goetz, Goetz, Gohlke, Golomb, Goncharov, González, Gopakumar, Gosselin, Gouaty, Gould, Grace, Grado, Granata, Granata, Grant, Gras, Grassia, Gray, Gray, Greco, Green, Green, Gretarsson, Gretarsson, Griffith, Griffiths, Griggs, Grignani, Grimaldi, Grimm, Grote, Grunewald, Gruning, Guerra, Guidi, Guimaraes, Guixé, Gulati, Guo, Guo, Gupta, Gupta, Gupta, Gustafson, Gustafson, Guzman, Ha, Haegel, Hagiwara, Haino, Halim, Hall, Hamilton, Hammond, Han, Haney, Hanks, Hanna, Hannam, Hannuksela, Hansen, Hansen, Hanson, Harder, Hardwick, Haris, Harms, Harry, Harry, Hartwig, Hasegawa, Haskell, Hasskew, Haster, Hattori, Haughian, Hayakawa, Hayama, Hayes, Healy, Heidmann, Heidt, Heintze, Heinze, Heinzel, Heitmann, Hellman, Hello, Helmling-Cornell, Hemming, Hendry, Heng, Hennes, Hennig, Hennig, Hernandez, Vivanco, Heurs, Hild, Hill, Himemoto, Hines, Hiranuma, Hirata, Hirose, Hochheim, Hofman, Hohmann, Holcomb, Holland, Hollows, Holmes, Holt, Holz, Hong, Hopkins, Hough, Hourihane, Howell, Hoy, Hoyland, Hreibi, Hsieh, Hsu, Huang, Huang, Huang, Huang, Huang, Huang, Hübner, Huddart, Hughey, Hui, Hui, Husa, Huttner, Huxford, Huynh-Dinh, Ide, Idzkowski, Iess, Ikenoue, Imam, Inayoshi, Ingram, Inoue, Ioka, Isi, Isleif, Ito, Itoh, Iyer, Izumi, JaberianHamedan, Jacqmin, Jadhav, Jadhav, James, Jan, Jani, Janquart, Janssens, Janthalur, Jaranowski, Jariwala, Jaume, Jenkins, Jenner, Jeon, Jeunon, Jia, Jin, Johns, Jones, Jones, Jones, Jones, Jones, Jonker, Ju, Jung, k.~Jung, Junker, Juste,
  Kaihotsu, Kajita, Kakizaki, Kalaghatgi, Kalogera, Kamai, Kamiizumi, Kanda, Kandhasamy, Kang, Kanner, Kao, Kapadia, Kapasi, Karat, Karathanasis, Karki, Kashyap, Kasprzack, Kastaun, Katsanevas, Katsavounidis, Katzman, Kaur, Kawabe, Kawaguchi, Kawai, Kawasaki, Kéfélian, Keitel, Key, Khadka, Khalili, Khan, Khazanov, Khetan, Khursheed, Kijbunchoo, Kim, Kim, Kim, Kim, Kim, Kim, Kimball, Kimura, Kinley-Hanlon, Kirchhoff, Kissel, Kita, Kitazawa, Kleybolte, Klimenko, Knee, Knowles, Knyazev, Koch, Koekoek, Kojima, Kokeyama, Koley, Kolitsidou, Kolstein, Komori, Kondrashov, Kong, Kontos, Koper, Korobko, Kotake, Kovalam, Kozak, Kozakai, Kozu, Kringel, Krishnendu, Królak, Kuehn, Kuei, Kuijer, Kumar, Kumar, Kumar, Kumar, Kume, Kuns, Kuo, Kuo, Kuromiya, Kuroyanagi, Kusayanagi, Kuwahara, Kwak, Lagabbe, Laghi, Lalande, Lam, Lamberts, Landry, Landry, Lane, Lang, Lange, Lantz, Rosa, Lartaux-Vollard, Lasky, Laxen, Lazzarini, Lazzaro, Leaci, Leavey, Lecoeuche, Lee, Lee, Lee, Lee, Lee, Lee, Lehmann, Lemaître, Leonardi, Leroy, Letendre, Levesque, Levin, Leviton, Leyde, Li, Li, Li, Li, Li, Li, Lin, Lin, Lin, Lin, Lin, Linde, Linker, Linley, Littenberg, Liu, Liu, Liu, Liu, Llamas, Llorens-Monteagudo, Lo, Lockwood, London, Longo, Lopez, Portilla, Lorenzini, Loriette, Lormand, Losurdo, Lott, Lough, Lousto, Lovelace, Lucaccioni, Lück, Lumaca, Lundgren, Luo, Lynam, Macas, MacInnis, Macleod, MacMillan, Macquet, Hernandez, Magazzù, Magee, Maggiore, Magnozzi, Mahesh, Majorana, Makarem, Maksimovic, Maliakal, Malik, Man, Mandic, Mangano, Mango, Mansell, Manske, Mantovani, Mapelli, Marchesoni, Marchio, Marion, Mark, Márka, Márka, Markakis, Markosyan, Markowitz, Maros, Marquina, Marsat, Martelli, Martin, Martin, Martinez, Martinez, Martinez, Martinovic, Martynov, Marx, Masalehdan, Mason, Massera, Masserot, Massinger, Masso-Reid, Mastrogiovanni, Matas, Mateu-Lucena, Matichard, Matiushechkina, Mavalvala, McCann, McCarthy, McClelland, McClincy, McCormick, McCuller, McGhee, McGuire, McIsaac,
  McIver, McRae, McWilliams, Meacher, Mehmet, Mehta, Meijer, Melatos, Melchor, Mendell, Menendez-Vazquez, Menoni, Mercer, Mereni, Merfeld, Merilh, Merritt, Merzougui, Meshkov, Messenger, Messick, Meyers, Meylahn, Mhaske, Miani, Miao, Michaloliakos, Michel, Michimura, Middleton, Milano, Miller, Miller, Miller, Miller, Millhouse, Mills, Milotti, Minazzoli, Minenkov, Mio, Mir, Miravet-Tenés, Mishra, Mishra, Mistry, Mitra, Mitrofanov, Mitselmakher, Mittleman, Miyakawa, Miyamoto, Miyazaki, Miyo, Miyoki, Mo, Moguel, Mogushi, Mohapatra, Mohite, Molina, Molina-Ruiz, Mondin, Montani, Moore, Moraru, Morawski, More, Moreno, Moreno, Mori, Morisaki, Moriwaki, Mours, Mow-Lowry, Mozzon, Muciaccia, Mukherjee, Mukherjee, Mukherjee, Mukherjee, Mukherjee, Mukund, Mullavey, Munch, Muñiz, Murray, Musenich, Muusse, Nadji, Nagano, Nagano, Nagar, Nakamura, Nakano, Nakano, Nakashima, Nakayama, Napolano, Nardecchia, Narikawa, Naticchioni, Nayak, Nayak, Negishi, Neil, Neilson, Nelemans, Nelson, Nery, Neubauer, Neunzert, Ng, Ng, Nguyen, Nguyen, Nguyen, Quynh, Ni, Nichols, Nishizawa, Nissanke, Nitoglia, Nocera, Norman, North, Nozaki, Nuttall, Oberling, O'Brien, Obuchi, O'Dell, Oelker, Ogaki, Oganesyan, Oh, Oh, Oh, Ohashi, Ohishi, Ohkawa, Ohme, Ohta, Okada, Okutani, Okutomi, Olivetto, Oohara, Ooi, Oram, O'Reilly, Ormiston, Ormsby, Ortega, O'Shaughnessy, O'Shea, Oshino, Ossokine, Osthelder, Otabe, Ottaway, Overmier, Pace, Pagano, Page, Pagliaroli, Pai, Pai, Palamos, Palashov, Palomba, Pan, Pan, Panda, Pang, Pang, Pankow, Pannarale, Pant, Panther, Paoletti, Paoli, Paolone, Parisi, Park, Park, Parker, Pascucci, Pasqualetti, Passaquieti, Passuello, Patel, Pathak, Patricelli, Patron, Paul, Payne, Pedraza, Pegoraro, Pele, Arellano, Penn, Perego, Pereira, Pereira, Perez, Périgois, Perkins, Perreca, Perriès, Petermann, Petterson, Pfeiffer, Pham, Phukon, Piccinni, Pichot, Piendibene, Piergiovanni, Pierini, Pierro, Pillant, Pillas, Pilo, Pinard, Pinto, Pinto, Piotrzkowski, Pirello, Pitkin, Placidi,
  Planas, Plastino, Pluchar, Poggiani, Polini, Pong, Ponrathnam, Popolizio, Porter, Poulton, Powell, Pracchia, Pradier, Prajapati, Prasai, Prasanna, Pratten, Principe, Prodi, Prokhorov, Prosposito, Prudenzi, Puecher, Punturo, Puosi, Puppo, Pürrer, Qi, Quetschke, Quitzow-James, Raab, Raaijmakers, Radkins, Radulesco, Raffai, Rail, Raja, Rajan, Ramirez, Ramirez, Ramos-Buades, Rana, Rapagnani, Rapol, Ray, Raymond, Raza, Razzano, Read, Rees, Regimbau, Rei, Reid, Reid, Reitze, Relton, Renzini, Rettegno, Rezac, Ricci, Richards, Richardson, Richardson, Riemenschneider, Riles, Rinaldi, Rink, Rizzo, Robertson, Robie, Robinet, Rocchi, Rodriguez, Rolland, Rollins, Romanelli, Romano, Romel, Romero-Rodríguez, Romero-Shaw, Romie, Ronchini, Rosa, Rose, Rosińska, Ross, Rowan, Rowlinson, Roy, Roy, Roy, Rozza, Ruggi, Ryan, Sachdev, Sadecki, Sadiq, Sago, Saito, Saito, Sakai, Sakai, Sakellariadou, Sakuno, Salafia, Salconi, Saleem, Salemi, Samajdar, Sanchez, Sanchez, Sanchez, Sanchis-Gual, Sanders, Sanuy, Saravanan, Sarin, Sassolas, Satari, Sathyaprakash, Sato, Sato, Sauter, Savage, Sawada, Sawant, Sawant, Sayah, Schaetzl, Scheel, Scheuer, Schiworski, Schmidt, Schmidt, Schnabel, Schneewind, Schofield, Schönbeck, Schulte, Schutz, Schwartz, Scott, Scott, Seglar-Arroyo, Sekiguchi, Sekiguchi, Sellers, Sengupta, Sentenac, Seo, Sequino, Sergeev, Setyawati, Shaffer, Shahriar, Shams, Shao, Sharma, Sharma, Shawhan, Shcheblanov, Shibagaki, Shikauchi, Shimizu, Shimoda, Shimode, Shinkai, Shishido, Shoda, Shoemaker, Shoemaker, ShyamSundar, Sieniawska, Sigg, Singer, Singh, Singh, Singha, Sintes, Sipala, Skliris, Slagmolen, Slaven-Blair, Smetana, Smith, Smith, Soldateschi, Somala, Somiya, Son, Soni, Soni, Sordini, Sorrentino, Sorrentino, Sotani, Soulard, Souradeep, Sowell, Spagnuolo, Spencer, Spera, Srinivasan, Srivastava, Srivastava, Staats, Stachie, Steer, Steinlechner, Steinlechner, Stops, Stover, Strain, Strang, Stratta, Strunk, Sturani, Stuver, Sudhagar, Sudhir, Sugimoto, Suh, Summerscales,
  Sun, Sun, Sunil, Sur, Suresh, Sutton, Suzuki, Suzuki, Swinkels, Szczepańczyk, Szewczyk, Tacca, Tagoshi, Tait, Takahashi, Takahashi, Takamori, Takano, Takeda, Takeda, Talbot, Talbot, Tanaka, Tanaka, Tanaka, Tanaka, Tanaka, Tanasijczuk, Tanioka, Tanner, Tao, Tao, Martín, Taranto, Tasson, Telada, Tenorio, Terhune, Terkowski, Thirugnanasambandam, Thomas, Thomas, Thompson, Thondapu, Thorne, Thrane, Tiwari, Tiwari, Tiwari, Toivonen, Toland, Tolley, Tomaru, Tomigami, Tomura, Tonelli, Torres-Forné, Torrie, e~Melo, Töyrä, Trapananti, Travasso, Traylor, Trevor, Tringali, Tripathee, Troiano, Trovato, Trozzo, Trudeau, Tsai, Tsai, Tsang, Tsang, Tsao, Tse, Tso, Tsubono, Tsuchida, Tsukada, Tsuna, Tsutsui, Tsuzuki, Turbang, Turconi, Tuyenbayev, Ubhi, Uchikata, Uchiyama, Udall, Ueda, Uehara, Ueno, Ueshima, Unnikrishnan, Uraguchi, Urban, Ushiba, Utina, Vahlbruch, Vajente, Vajpeyi, Valdes, Valentini, Valsan, van Bakel, van Beuzekom, van~den Brand, Broeck, Vander-Hyde, van~der Schaaf, van Heijningen, Vanosky, van Putten, van Remortel, Vardaro, Vargas, Varma, Vasúth, Vecchio, Vedovato, Veitch, Veitch, Venneberg, Venugopalan, Verkindt, Verma, Verma, Veske, Vetrano, Viceré, Vidyant, Viets, Vijaykumar, Villa-Ortega, Vinet, Virtuoso, Vitale, Vo, Vocca, von Reis, von Wrangel, Vorvick, Vyatchanin, Wade, Wade, Wagner, Walet, Walker, Wallace, Wallace, Walsh, Wang, Wang, Wang, Ward, Warner, Was, Washimi, Washington, Watchi, Weaver, Webster, Weinert, Weinstein, Weiss, Weller, Wellmann, Wen, Weßels, Wette, Whelan, White, Whiting, Whittle, Wilken, Williams, Williams, Williamson, Willis, Willke, Wilson, Winkler, Wipf, Wlodarczyk, Woan, Woehler, Wofford, Wong, Wu, Wu, Wu, Wu, Wysocki, Xiao, Xu, Yamada, Yamamoto, Yamamoto, Yamamoto, Yamamoto, Yamashita, Yamazaki, Yang, Yang, Yang, Yang, Yang, Yap, Yeeles, Yelikar, Ying, Yokogawa, Yokoyama, Yokozawa, Yoo, Yoshioka, Yu, Yu, Yuzurihara, Zadrożny, Zanolin, Zeidler, Zelenova, Zendri, Zevin, Zhan, Zhang, Zhang, Zhang, Zhang, Zhang, Zhao, Zhao,
  Zhao, Zhao, Zhou, Zhou, Zhu, Zhu, Zimmerman, Zlochower, Zucker, \& Zweizig}]{o3_rates_and_pops}
Abbott, R., Abbott, T.~D., Acernese, F., {et~al.} 2022, The population of merging compact binaries inferred using gravitational waves through GWTC-3.
\newblock \doarXiv{2111.03634}

\bibitem[{Acernese {et~al.}(2015)}]{advancedvirgo}
Acernese, F., {et~al.} 2015, Class. Quant. Grav., 32, 024001, \dodoi{10.1088/0264-9381/32/2/024001}

\bibitem[{Akutsu {et~al.}(2021)Akutsu, Ando, Arai, Arai, Araki, Araya, Aritomi, Asada, Aso, Bae, Bae, Baiotti, Bajpai, Barton, Cannon, Cao, Capocasa, Chan, Chen, Chen, Chen, Chiang, Chu, Chu, Eguchi, Enomoto, Flaminio, Fujii, Fujikawa, Fukunaga, Fukushima, Gao, Ge, Ha, Hagiwara, Haino, Han, Hasegawa, Hattori, Hayakawa, Hayama, Himemoto, Hiranuma, Hirata, Hirose, Hong, Hsieh, Huang, Huang, Huang, Huang, Huang, Hui, Ide, Ikenoue, Imam, Inayoshi, Inoue, Ioka, Ito, Itoh, Izumi, Jeon, Jin, Jung, Jung, Kaihotsu, Kajita, Kakizaki, Kamiizumi, Kanda, Kang, Kawaguchi, Kawai, Kawasaki, Kim, Kim, Kim, Kim, Kim, Kimura, Kita, Kitazawa, Kojima, Kokeyama, Komori, Kong, Kotake, Kozakai, Kozu, Kumar, Kume, Kuo, Kuo, Kuromiya, Kuroyanagi, Kusayanagi, Kwak, Lee, Lee, Lee, Leonardi, Li, Lin, Lin, Lin, Lin, Lin, Liu, Luo, Majorana, Marchio, Michimura, Mio, Miyakawa, Miyamoto, Miyazaki, Miyo, Miyoki, Mori, Morisaki, Moriwaki, Nagano, Nagano, Nakamura, Nakano, Nakano, Nakashima, Nakayama, Narikawa, Naticchioni, Negishi, Nguyen~Quynh, Ni, Nishizawa, Nozaki, Obuchi, Ogaki, Oh, Oh, Oh, Ohashi, Ohishi, Ohkawa, Ohta, Okutani, Okutomi, Oohara, Ooi, Oshino, Otabe, Pan, Pang, Parisi, Park, Peña~Arellano, Pinto, Sago, Saito, Saito, Sakai, Sakai, Sakuno, Sato, Sato, Sawada, Sekiguchi, Sekiguchi, Shao, Shibagaki, Shimizu, Shimoda, Shimode, Shinkai, Shishido, Shoda, Somiya, Son, Sotani, Sugimoto, Suresh, Suzuki, Suzuki, Tagoshi, Takahashi, Takahashi, Takamori, Takano, Takeda, Takeda, Tanaka, Tanaka, Tanaka, Tanaka, Tanaka, Tanioka, Tapia San~Martin, Telada, Tomaru, Tomigami, Tomura, Travasso, Trozzo, Tsang, Tsao, Tsubono, Tsuchida, Tsutsui, Tsuzuki, Tuyenbayev, Uchikata, Uchiyama, Ueda, Uehara, Ueno, Ueshima, Uraguchi, Ushiba, van Putten, Vocca, Wang, Washimi, Wu, Wu, Wu, Xu, Yamada, Yamamoto, Yamamoto, Yamamoto, Yamashita, Yamazaki, Yang, Yokogawa, Yokoyama, Yokozawa, Yoshioka, Yuzurihara, Zeidler, Zhan, Zhang, Zhao, \& Zhu}]{kagra_overview}
Akutsu, T., Ando, M., Arai, K., {et~al.} 2021, Progress of Theoretical and Experimental Physics, 2021, \dodoi{10.1093/ptep/ptab018}

\bibitem[{Aso {et~al.}(2013)Aso, Michimura, Somiya, Ando, Miyakawa, Sekiguchi, Tatsumi, \& Yamamoto}]{kagra_design}
Aso, Y., Michimura, Y., Somiya, K., {et~al.} 2013, Phys. Rev. D, 88, 043007, \dodoi{10.1103/PhysRevD.88.043007}

\bibitem[{Binney \& Tremaine(1987)}]{binney_tremaine_galactic_dynamics}
Binney, J., \& Tremaine, S. 1987, Galactic Dynamics, Princeton Series in Astrophysics (Princeton University Press)

\bibitem[{Cao {et~al.}(2014)Cao, Li, \& Wang}]{cao_lensing}
Cao, Z., Li, L.-F., \& Wang, Y. 2014, Phys. Rev. D, 90, 062003, \dodoi{10.1103/PhysRevD.90.062003}

\bibitem[{Dai \& Venumadhav(2017)}]{dai_lensed_waveforms}
Dai, L., \& Venumadhav, T. 2017, On the waveforms of gravitationally lensed gravitational waves,  arXiv, \dodoi{10.48550/ARXIV.1702.04724}

\bibitem[{Dai {et~al.}(2017)Dai, Venumadhav, \& Sigurdson}]{dai_galaxy_lensing}
Dai, L., Venumadhav, T., \& Sigurdson, K. 2017, Phys. Rev. D, 95, 044011, \dodoi{10.1103/PhysRevD.95.044011}

\bibitem[{Dyson {et~al.}(1920)Dyson, Eddington, \& Davidson}]{eddington_experiment}
Dyson, F.~W., Eddington, A.~S., \& Davidson, C. 1920, Philosophical Transactions of the Royal Society of London. Series A, Containing Papers of a Mathematical or Physical Character, 220, 291, \dodoi{10.1098/rsta.1920.0009}

\bibitem[{Einstein(1936)}]{Einstein_lensing}
Einstein, A. 1936, Science, 84, 506, \dodoi{10.1126/science.84.2188.506}

\bibitem[{Ezquiaga {et~al.}(2021)Ezquiaga, Holz, Hu, Lagos, \& Wald}]{ezquiaga_phase_effects}
Ezquiaga, J.~M., Holz, D.~E., Hu, W., Lagos, M., \& Wald, R.~M. 2021, Phys. Rev. D, 103, 064047, \dodoi{10.1103/PhysRevD.103.064047}

\bibitem[{Ezquiaga \& Zumalac\'arregui(2020)}]{ezquiaga_testing_gr}
Ezquiaga, J.~M., \& Zumalac\'arregui, M. 2020, Phys. Rev. D, 102, 124048, \dodoi{10.1103/PhysRevD.102.124048}

\bibitem[{Fan {et~al.}(2017)Fan, Liao, Biesiada, Pi\'orkowska-Kurpas, \& Zhu}]{fan_speed_of_gravity}
Fan, X.-L., Liao, K., Biesiada, M., Pi\'orkowska-Kurpas, A., \& Zhu, Z.-H. 2017, Phys. Rev. Lett., 118, 091102, \dodoi{10.1103/PhysRevLett.118.091102}

\bibitem[{Goyal {et~al.}(2021)Goyal, Haris, Mehta, \& Ajith}]{goyal_testing_gr}
Goyal, S., Haris, K., Mehta, A.~K., \& Ajith, P. 2021, Phys. Rev. D, 103, 024038, \dodoi{10.1103/PhysRevD.103.024038}

\bibitem[{Hannuksela {et~al.}(2020)Hannuksela, Collett, Çalışkan, \& Li}]{hannuksela_lensing_localisation}
Hannuksela, O.~A., Collett, T.~E., Çalışkan, M., \& Li, T. G.~F. 2020, Monthly Notices of the Royal Astronomical Society, 498, 3395, \dodoi{10.1093/mnras/staa2577}

\bibitem[{Hannuksela {et~al.}(2019)Hannuksela, Haris, Ng, Kumar, Mehta, Keitel, Li, \& Ajith}]{o2_lensing}
Hannuksela, O.~A., Haris, K., Ng, K. K.~Y., {et~al.} 2019, The Astrophysical Journal Letters, 874, L2, \dodoi{10.3847/2041-8213/ab0c0f}

\bibitem[{{Haris} {et~al.}(2018){Haris}, {Mehta}, {Kumar}, {Venumadhav}, \& {Ajith}}]{harris_posterior_overlap}
{Haris}, K., {Mehta}, A.~K., {Kumar}, S., {Venumadhav}, T., \& {Ajith}, P. 2018, arXiv e-prints, arXiv:1807.07062, \dodoi{10.48550/arXiv.1807.07062}

\bibitem[{Haris {et~al.}(2018)Haris, Mehta, Kumar, Venumadhav, \& Ajith}]{haris_mgal}
Haris, K., Mehta, A.~K., Kumar, S., Venumadhav, T., \& Ajith, P. 2018, {Identifying strongly lensed gravitational wave signals from binary black hole mergers}.
\newblock \doarXiv{1807.07062}

\bibitem[{Herrera-Martín {et~al.}(2019)Herrera-Martín, Hendry, Gonzalez-Morales, \& Ureña-López}]{antonio_lens_modelling}
Herrera-Martín, A., Hendry, M., Gonzalez-Morales, A.~X., \& Ureña-López, L.~A. 2019, The Astrophysical Journal, 872, 11, \dodoi{10.3847/1538-4357/aafaf0}

\bibitem[{Janquart {et~al.}(2021)Janquart, Hannuksela, Haris, \& Van Den Broeck}]{golum_paper}
Janquart, J., Hannuksela, O.~A., Haris, K., \& Van Den Broeck, C. 2021, Monthly Notices of the Royal Astronomical Society, 506, 5430, \dodoi{10.1093/mnras/stab1991}

\bibitem[{Janquart {et~al.}(2022{\natexlab{a}})Janquart, Haris, \& Hannuksela}]{git_golum}
Janquart, J., Haris, K., \& Hannuksela, O. 2022{\natexlab{a}}, {GOLUM: a software for rapid strongly-lensed gravitational wave parameter estimation }, \url{https://github.com/lemnis12/golum}

\bibitem[{Janquart {et~al.}(2023{\natexlab{a}})Janquart, Haris, Hannuksela, \& Van Den~Broeck}]{golum_update_paper}
Janquart, J., Haris, K., Hannuksela, O.~A., \& Van Den~Broeck, C. 2023{\natexlab{a}}, {The Return of GOLUM: Improving Distributed Joint Parameter Estimation for Strongly-Lensed Gravitational Waves}.
\newblock \doarXiv{2304.12148}

\bibitem[{Janquart {et~al.}(2022{\natexlab{b}})Janquart, More, \& Van Den~Broeck}]{janquart_catalog_selection}
Janquart, J., More, A., \& Van Den~Broeck, C. 2022{\natexlab{b}}, Mon. Not. Roy. Astron. Soc., 519, 2046, \dodoi{10.1093/mnras/stac3660}

\bibitem[{Janquart {et~al.}(2023{\natexlab{b}})Janquart, Wright, Goyal, Chan, Ganguly, Ángel Garrón, Keitel, Li, Liu, Lo, Mishra, More, Phurailatpam, Pankunni, Biscoveanu, Cremonese, Cudell, Ezquiaga, Garcia-Bellido, Hannuksela, Haris, Harry, Hendry, Husa, Kapadia, Li, Hernandez, Mukherjee, Seo, Broek, \& Veitch}]{o3_technical}
Janquart, J., Wright, M., Goyal, S., {et~al.} 2023{\natexlab{b}}, Follow-up analyses to the O3 LIGO-Virgo-KAGRA lensing searches.
\newblock \doarXiv{2306.03827}

\bibitem[{Li {et~al.}(2018)Li, Mao, Zhao, \& Lu}]{li_lensing_rate}
Li, S.-S., Mao, S., Zhao, Y., \& Lu, Y. 2018, Monthly Notices of the Royal Astronomical Society, 476, 2220, \dodoi{10.1093/mnras/sty411}

\bibitem[{{Liao} {et~al.}(2017){Liao}, {Fan}, {Ding}, {Biesiada}, \& {Zhu}}]{liao_cosmography}
{Liao}, K., {Fan}, X.-L., {Ding}, X., {Biesiada}, M., \& {Zhu}, Z.-H. 2017, Nature Communications, 8, 1148, \dodoi{10.1038/s41467-017-01152-9}

\bibitem[{{LIGO Scientific Collaboration and Virgo Collaboration and KAGRA Collaboration}(2023)}]{GWTC3_pop_dataRelease}
{LIGO Scientific Collaboration and Virgo Collaboration and KAGRA Collaboration}. 2023, {The population of merging compact binaries inferred using gravitational waves through GWTC-3 - Data release (Version v2)}, {\url{https://zenodo.org/record/5655785}}

\bibitem[{{Liu} {et~al.}(2021){Liu}, {Maga{\~n}a Hernandez}, \& {Creighton}}]{liu_selection_effects}
{Liu}, X., {Maga{\~n}a Hernandez}, I., \& {Creighton}, J. 2021, \apj, 908, 97, \dodoi{10.3847/1538-4357/abd7eb}

\bibitem[{{Lo} \& {Magana Hernandez}(2021)}]{hanabi_paper}
{Lo}, R. K.~L., \& {Magana Hernandez}, I. 2021, arXiv e-prints, arXiv:2104.09339.
\newblock \doarXiv{2104.09339}

\bibitem[{Mishra {et~al.}(2021)Mishra, Meena, More, Bose, \& Bagla}]{mishra_microlensing_population}
Mishra, A., Meena, A.~K., More, A., Bose, S., \& Bagla, J.~S. 2021, Monthly Notices of the Royal Astronomical Society, 508, 4869, \dodoi{10.1093/mnras/stab2875}

\bibitem[{More \& More(2022)}]{more_mgal}
More, A., \& More, S. 2022, Mon. Not. Roy. Astron. Soc., 515, 1044, \dodoi{10.1093/mnras/stac1704}

\bibitem[{Nakamura \& Deguchi(1999)}]{nakamura_review_lensing}
Nakamura, T.~T., \& Deguchi, S. 1999, Progress of Theoretical Physics Supplement, 133, 137, \dodoi{10.1143/PTPS.133.137}

\bibitem[{Ng {et~al.}(2018)Ng, Wong, Broadhurst, \& Li}]{ng_rate_galaxy_lensing}
Ng, K. K.~Y., Wong, K. W.~K., Broadhurst, T., \& Li, T. G.~F. 2018, Phys. Rev. D, 97, 023012, \dodoi{10.1103/PhysRevD.97.023012}

\bibitem[{{Ohanian}(1974)}]{focusing_gravitational_radiation}
{Ohanian}, H.~C. 1974, International Journal of Theoretical Physics, 9, 425, \dodoi{10.1007/BF01810927}

\bibitem[{Pratten {et~al.}(2021)}]{IMRPhenomXPHM_paper}
Pratten, G., {et~al.} 2021, Phys. Rev. D, 103, 104056, \dodoi{10.1103/PhysRevD.103.104056}

\bibitem[{Robertson {et~al.}(2020)Robertson, Smith, Massey, Eke, Jauzac, Bianconi, \& Ryczanowski}]{robertson_strong_lensing}
Robertson, A., Smith, G.~P., Massey, R., {et~al.} 2020, Monthly Notices of the Royal Astronomical Society, 495, 3727, \dodoi{10.1093/mnras/staa1429}

\bibitem[{{Schneider} {et~al.}(1992){Schneider}, {Ehlers}, \& {Falco}}]{schneider_gravitational_lenses}
{Schneider}, P., {Ehlers}, J., \& {Falco}, E.~E. 1992, {Gravitational Lenses} (Springer Science \& Business Media), \dodoi{10.1007/978-3-662-03758-4}

\bibitem[{Sereno {et~al.}(2011)Sereno, Jetzer, Sesana, \& Volonteri}]{sereno_lensing_cosmography}
Sereno, M., Jetzer, P., Sesana, A., \& Volonteri, M. 2011, Monthly Notices of the Royal Astronomical Society, 415, 2773, \dodoi{10.1111/j.1365-2966.2011.18895.x}

\bibitem[{Smith {et~al.}(2018)Smith, Jauzac, Veitch, Farr, Massey, \& Richard}]{smith_cluster_lensing_mnras}
Smith, G.~P., Jauzac, M., Veitch, J., {et~al.} 2018, Monthly Notices of the Royal Astronomical Society, 475, 3823, \dodoi{10.1093/mnras/sty031}

\bibitem[{Smith {et~al.}(2017)Smith, Berry, Bianconi, Farr, Jauzac, Massey, Richard, Robertson, Sharon, Vecchio, \& et~al.}]{smith_cluster_lensing}
Smith, G.~P., Berry, C., Bianconi, M., {et~al.} 2017, Proceedings of the International Astronomical Union, 13, 98–102, \dodoi{10.1017/S1743921318003757}

\bibitem[{Smith {et~al.}(2023)Smith, Robertson, Mahler, Nicholl, Ryczanowski, Bianconi, Sharon, Massey, Richard, \& Jauzac}]{smith_rate_2023}
Smith, G.~P., Robertson, A., Mahler, G., {et~al.} 2023, Monthly Notices of the Royal Astronomical Society, 520, 702, \dodoi{10.1093/mnras/stad140}

\bibitem[{Somiya(2012)}]{kagra_configuration}
Somiya, K. 2012, Classical and Quantum Gravity, 29, 124007, \dodoi{10.1088/0264-9381/29/12/124007}

\bibitem[{Takahashi \& Nakamura(2003)}]{takahashi_nakamura_lensing}
Takahashi, R., \& Nakamura, T. 2003, The Astrophysical Journal, 595, 1039, \dodoi{10.1086/377430}

\bibitem[{Tambalo {et~al.}(2022)Tambalo, Zumalacárregui, Dai, \& Cheung}]{tambalo_lenses}
Tambalo, G., Zumalacárregui, M., Dai, L., \& Cheung, M. H.-Y. 2022, Gravitational wave lensing as a probe of halo properties and dark matter.
\newblock \doarXiv{2212.11960}

\bibitem[{{The LIGO Scientific Collaboration} {et~al.}(2021){The LIGO Scientific Collaboration}, {The Virgo Collaboration}, {The KAGRA Collaboration}, Abbott, Abbott, Acernese, Ackley, Adams, Adhikari, Adhikari, Adya, Affeldt, Agarwal, Agathos, Agatsuma, Aggarwal, Aguiar, Aiello, Ain, Ajith, Akcay, Akutsu, Albanesi, Allocca, Altin, Amato, Anand, Anand, Ananyeva, Anderson, Anderson, Ando, Andrade, Andres, Andrić, Angelova, Ansoldi, Antelis, Antier, Appert, Arai, Arai, Arai, Araki, Araya, Araya, Areeda, Arène, Aritomi, Arnaud, Arogeti, Aronson, Arun, Asada, Asali, Ashton, Aso, Assiduo, Aston, Astone, Aubin, Austin, Babak, Badaracco, Bader, Badger, Bae, Bae, Baer, Bagnasco, Bai, Baiotti, Baird, Bajpai, Ball, Ballardin, Ballmer, Balsamo, Baltus, Banagiri, Bankar, Barayoga, Barbieri, Barish, Barker, Barneo, Barone, Barr, Barsotti, Barsuglia, Barta, Bartlett, Barton, Bartos, Bassiri, Basti, Bawaj, Bayley, Baylor, Bazzan, Bécsy, Bedakihale, Bejger, Belahcene, Benedetto, Beniwal, Bennett, Bentley, BenYaala, Bergamin, Berger, Bernuzzi, Berry, Bersanetti, Bertolini, Betzwieser, Beveridge, Bhandare, Bhardwaj, Bhattacharjee, Bhaumik, Bilenko, Billingsley, Bini, Birney, Birnholtz, Biscans, Bischi, Biscoveanu, Bisht, Biswas, Bitossi, Bizouard, Blackburn, Blair, Blair, Blair, Bobba, Bode, Boer, Bogaert, Boldrini, Bonavena, Bondu, Bonilla, Bonnand, Booker, Boom, Bork, Boschi, Bose, Bose, Bossilkov, Boudart, Bouffanais, Bozzi, Bradaschia, Brady, Bramley, Branch, Branchesi, Brandt, Brau, Breschi, Briant, Briggs, Brillet, Brinkmann, Brockill, Brooks, Brooks, Brown, Brunett, Bruno, Bruntz, Bryant, Bulik, Bulten, Buonanno, Buscicchio, Buskulic, Buy, Byer, Davies, Cadonati, Cagnoli, Cahillane, Bustillo, Callaghan, Callister, Calloni, Cameron, Camp, Canepa, Canevarolo, Cannavacciuolo, Cannon, Cao, Cao, Capocasa, Capote, Carapella, Carbognani, Carlin, Carney, Carpinelli, Carrillo, Carullo, Carver, Diaz, Casentini, Castaldi, Caudill, Cavaglià, Cavalier, Cavalieri, Ceasar, Cella,
  Cerdá-Durán, Cesarini, Chaibi, Chakravarti, Subrahmanya, Champion, Chan, Chan, Chan, Chan, Chan, Chandra, Chanial, Chao, Chapman-Bird, Charlton, Chase, Chassande-Mottin, Chatterjee, Chatterjee, Chatterjee, Chaturvedi, Chaty, Chatziioannou, Chen, Chen, Chen, Chen, Chen, Chen, Chen, Chen, Cheng, Cheong, Cheung, Chia, Chiadini, Chiang, Chiarini, Chierici, Chincarini, Chiofalo, Chiummo, Cho, Cho, Choudhary, Choudhary, Christensen, Chu, Chu, Chu, Chua, Chung, Ciani, Ciecielag, Cieślar, Cifaldi, Ciobanu, Ciolfi, Cipriano, Cirone, Clara, Clark, Clark, Clarke, Clearwater, Clesse, Cleva, Coccia, Codazzo, Cohadon, Cohen, Cohen, Colleoni, Collette, Colombo, Colpi, Compton, Constancio, Conti, Cooper, Corban, Corbitt, Cordero-Carrión, Corezzi, Corley, Cornish, Corre, Corsi, Cortese, Costa, Cotesta, Coughlin, Coulon, Countryman, Cousins, Couvares, Coward, Cowart, Coyne, Coyne, Creighton, Creighton, Criswell, Croquette, Crowder, Cudell, Cullen, Cumming, Cummings, Cunningham, Cuoco, Curyło, Dabadie, Canton, Dall'Osso, Dálya, Dana, DaneshgaranBajastani, D'Angelo, Danila, Danilishin, D'Antonio, Danzmann, Darsow-Fromm, Dasgupta, Datrier, Datta, Dattilo, Dave, Davier, Davis, Davis, Daw, de~Alarcón, Dean, DeBra, Deenadayalan, Degallaix, De~Laurentis, Deléglise, Del~Favero, De~Lillo, De~Lillo, Del~Pozzo, DeMarchi, De~Matteis, D'Emilio, Demos, Dent, Depasse, De~Pietri, De~Rosa, De~Rossi, DeSalvo, De~Simone, Dhurandhar, Díaz, Diaz-Ortiz, Didio, Dietrich, Di~Fiore, Di~Fronzo, Di~Giorgio, Di~Giovanni, Di~Giovanni, Di~Girolamo, Di~Lieto, Ding, Di~Pace, Di~Palma, Di~Renzo, Divakarla, Dmitriev, Doctor, D'Onofrio, Donovan, Dooley, Doravari, Dorrington, Drago, Driggers, Drori, Ducoin, Dupej, Durante, D'Urso, Duverne, Dwyer, Eassa, Easter, Ebersold, Eckhardt, Eddolls, Edelman, Edo, Edy, Effler, Eguchi, Eichholz, Eikenberry, Eisenmann, Eisenstein, Ejlli, Engelby, Enomoto, Errico, Essick, Estellés, Estevez, Etienne, Etzel, Evans, Evans, Ewing, Fafone, Fair, Fairhurst, Farah, Farinon,
  Farr, Farr, Farrow, Fauchon-Jones, Favaro, Favata, Fays, Fazio, Feicht, Fejer, Fenyvesi, Ferguson, Fernandez-Galiana, Ferrante, Ferreira, Fidecaro, Figura, Fiori, Fishbach, Fisher, Fittipaldi, Fiumara, Flaminio, Floden, Fong, Font, Fornal, Forsyth, Franke, Frasca, Frasconi, Frederick, Freed, Frei, Freise, Frey, Fritschel, Frolov, Fronzé, Fujii, Fujikawa, Fukunaga, Fukushima, Fulda, Fyffe, Gabbard, Gabella, Gadre, Gair, Gais, Galaudage, Gamba, Ganapathy, Ganguly, Gao, Gaonkar, Garaventa, García, García-Núñez, García-Quirós, Garufi, Gateley, Gaudio, Gayathri, Ge, Gemme, Gennai, George, George, Gerberding, Gergely, Gewecke, Ghonge, Ghosh, Ghosh, Ghosh, Ghosh, Giacomazzo, Giacoppo, Giaime, Giardina, Gibson, Gier, Giesler, Giri, Gissi, Glanzer, Gleckl, Godwin, Goetz, Goetz, Gohlke, Golomb, Goncharov, González, Gopakumar, Gosselin, Gouaty, Gould, Grace, Grado, Granata, Granata, Grant, Gras, Grassia, Gray, Gray, Greco, Green, Green, Gretarsson, Gretarsson, Griffith, Griffiths, Griggs, Grignani, Grimaldi, Grimm, Grote, Grunewald, Gruning, Guerra, Guidi, Guimaraes, Guixé, Gulati, Guo, Guo, Gupta, Gupta, Gupta, Gustafson, Gustafson, Guzman, Ha, Haegel, Hagiwara, Haino, Halim, Hall, Hamilton, Hammond, Han, Haney, Hanks, Hanna, Hannam, Hannuksela, Hansen, Hansen, Hanson, Harder, Hardwick, Haris, Harms, Harry, Harry, Hartwig, Hasegawa, Haskell, Hasskew, Haster, Hattori, Haughian, Hayakawa, Hayama, Hayes, Healy, Heidmann, Heidt, Heintze, Heinze, Heinzel, Heitmann, Hellman, Hello, Helmling-Cornell, Hemming, Hendry, Heng, Hennes, Hennig, Hennig, Hernandez, Vivanco, Heurs, Hild, Hill, Himemoto, Hines, Hiranuma, Hirata, Hirose, Hochheim, Hofman, Hohmann, Holcomb, Holland, Holley-Bockelmann, Hollows, Holmes, Holt, Holz, Hong, Hopkins, Hough, Hourihane, Howell, Hoy, Hoyland, Hreibi, Hsieh, Hsu, Huang, Huang, Huang, Huang, Huang, Huang, Hübner, Huddart, Hughey, Hui, Hui, Husa, Huttner, Huxford, Huynh-Dinh, Ide, Idzkowski, Iess, Ikenoue, Imam, Inayoshi, Ingram, Inoue, Ioka,
  Isi, Isleif, Ito, Itoh, Iyer, Izumi, JaberianHamedan, Jacqmin, Jadhav, Jadhav, James, Jan, Jani, Janquart, Janssens, Janthalur, Jaranowski, Jariwala, Jaume, Jenkins, Jenner, Jeon, Jeunon, Jia, Jin, Johns, Johnson-McDaniel, Jones, Jones, Jones, Jones, Jones, Jonker, Ju, Jung, Jung, Junker, Juste, Kaihotsu, Kajita, Kakizaki, Kalaghatgi, Kalogera, Kamai, Kamiizumi, Kanda, Kandhasamy, Kang, Kanner, Kao, Kapadia, Kapasi, Karat, Karathanasis, Karki, Kashyap, Kasprzack, Kastaun, Katsanevas, Katsavounidis, Katzman, Kaur, Kawabe, Kawaguchi, Kawai, Kawasaki, Kéfélian, Keitel, Key, Khadka, Khalili, Khan, Khazanov, Khetan, Khursheed, Kijbunchoo, Kim, Kim, Kim, Kim, Kim, Kim, Kimball, Kimura, Kinley-Hanlon, Kirchhoff, Kissel, Kita, Kitazawa, Kleybolte, Klimenko, Knee, Knowles, Knyazev, Koch, Koekoek, Kojima, Kokeyama, Koley, Kolitsidou, Kolstein, Komori, Kondrashov, Kong, Kontos, Koper, Korobko, Kotake, Kovalam, Kozak, Kozakai, Kozu, Kringel, Krishnendu, Królak, Kuehn, Kuei, Kuijer, Kulkarni, Kumar, Kumar, Kumar, Kumar, Kume, Kuns, Kuo, Kuo, Kuromiya, Kuroyanagi, Kusayanagi, Kuwahara, Kwak, Lagabbe, Laghi, Lalande, Lam, Lamberts, Landry, Lane, Lang, Lange, Lantz, La~Rosa, Lartaux-Vollard, Lasky, Laxen, Lazzarini, Lazzaro, Leaci, Leavey, Lecoeuche, Lee, Lee, Lee, Lee, Lee, Lee, Lehmann, Lemaître, Leonardi, Leroy, Letendre, Levesque, Levin, Leviton, Leyde, Li, Li, Li, Li, Li, Li, Lin, Lin, Lin, Lin, Lin, Linde, Linker, Linley, Littenberg, Liu, Liu, Liu, Liu, Llamas, Llorens-Monteagudo, Lo, Lockwood, Loh, London, Longo, Lopez, Portilla, Lorenzini, Loriette, Lormand, Losurdo, Lott, Lough, Lousto, Lovelace, Lucaccioni, Lück, Lumaca, Lundgren, Luo, Lynam, Macas, MacInnis, Macleod, MacMillan, Macquet, Hernandez, Magazzù, Magee, Maggiore, Magnozzi, Mahesh, Majorana, Makarem, Maksimovic, Maliakal, Malik, Man, Mandic, Mangano, Mango, Mansell, Manske, Mantovani, Mapelli, Marchesoni, Marchio, Marion, Mark, Márka, Márka, Markakis, Markosyan, Markowitz, Maros, Marquina, Marsat,
  Martelli, Martin, Martin, Martinez, Martinez, Martinez, Martinovic, Martynov, Marx, Masalehdan, Mason, Massera, Masserot, Massinger, Masso-Reid, Mastrogiovanni, Matas, Mateu-Lucena, Matichard, Matiushechkina, Mavalvala, McCann, McCarthy, McClelland, McClincy, McCormick, McCuller, McGhee, McGuire, McIsaac, McIver, McRae, McWilliams, Meacher, Mehmet, Mehta, Meijer, Melatos, Melchor, Mendell, Menendez-Vazquez, Menoni, Mercer, Mereni, Merfeld, Merilh, Merritt, Merzougui, Meshkov, Messenger, Messick, Meyers, Meylahn, Mhaske, Miani, Miao, Michaloliakos, Michel, Michimura, Middleton, Milano, Miller, Miller, Miller, Millhouse, Mills, Milotti, Minazzoli, Minenkov, Mio, Mir, Miravet-Tenés, Mishra, Mishra, Mistry, Mitra, Mitrofanov, Mitselmakher, Mittleman, Miyakawa, Miyamoto, Miyazaki, Miyo, Miyoki, Mo, Modafferi, Moguel, Mogushi, Mohapatra, Mohite, Molina, Molina-Ruiz, Mondin, Montani, Moore, Moraru, Morawski, More, Moreno, Moreno, Mori, Morisaki, Moriwaki, Morrás, Mours, Mow-Lowry, Mozzon, Muciaccia, Mukherjee, Mukherjee, Mukherjee, Mukherjee, Mukherjee, Mukund, Mullavey, Munch, Muñiz, Murray, Musenich, Muusse, Nadji, Nagano, Nagano, Nagar, Nakamura, Nakano, Nakano, Nakashima, Nakayama, Napolano, Nardecchia, Narikawa, Naticchioni, Nayak, Nayak, Negishi, Neil, Neilson, Nelemans, Nelson, Nery, Neubauer, Neunzert, Ng, Ng, Nguyen, Nguyen, Nguyen, Quynh, Ni, Nichols, Nishizawa, Nissanke, Nitoglia, Nocera, Norman, North, Nozaki, Siles, Nuttall, Oberling, O'Brien, Obuchi, O'Dell, Oelker, Ogaki, Oganesyan, Oh, Oh, Oh, Ohashi, Ohishi, Ohkawa, Ohme, Ohta, Okada, Okutani, Okutomi, Olivetto, Oohara, Ooi, Oram, O'Reilly, Ormiston, Ormsby, Ortega, O'Shaughnessy, O'Shea, Oshino, Ossokine, Osthelder, Otabe, Ottaway, Overmier, Pace, Pagano, Page, Pagliaroli, Pai, Pai, Palamos, Palashov, Palomba, Pan, Pan, Panda, Pang, Pang, Pankow, Pannarale, Pant, Panther, Paoletti, Paoli, Paolone, Parisi, Park, Park, Parker, Pascucci, Pasqualetti, Passaquieti, Passuello, Patel, Pathak, Patricelli,
  Patron, Paul, Payne, Pedraza, Pegoraro, Pele, Arellano, Penn, Perego, Pereira, Pereira, Perez, Périgois, Perkins, Perreca, Perriès, Petermann, Petterson, Pfeiffer, Pham, Phukon, Piccinni, Pichot, Piendibene, Piergiovanni, Pierini, Pierro, Pillant, Pillas, Pilo, Pinard, Pinto, Pinto, Piotrzkowski, Piotrzkowski, Pirello, Pitkin, Placidi, Planas, Plastino, Pluchar, Poggiani, Polini, Pong, Ponrathnam, Popolizio, Porter, Poulton, Powell, Pracchia, Pradier, Prajapati, Prasai, Prasanna, Pratten, Principe, Prodi, Prokhorov, Prosposito, Prudenzi, Puecher, Punturo, Puosi, Puppo, Pürrer, Qi, Quetschke, Quitzow-James, Qutob, Raab, Raaijmakers, Radkins, Radulesco, Raffai, Rail, Raja, Rajan, Ramirez, Ramirez, Ramos-Buades, Rana, Rapagnani, Rapol, Ray, Raymond, Raza, Razzano, Read, Rees, Regimbau, Rei, Reid, Reid, Reitze, Relton, Renzini, Rettegno, Reza, Rezac, Ricci, Richards, Richardson, Richardson, Riemenschneider, Riles, Rinaldi, Rink, Rizzo, Robertson, Robie, Robinet, Rocchi, Rodriguez, Rolland, Rollins, Romanelli, Romano, Romel, Romero-Rodríguez, Romero-Shaw, Romie, Ronchini, Rosa, Rose, Rosińska, Ross, Rowan, Rowlinson, Roy, Roy, Roy, Rozza, Ruggi, Ruiz-Rocha, Ryan, Sachdev, Sadecki, Sadiq, Sago, Saito, Saito, Sakai, Sakai, Sakellariadou, Sakuno, Salafia, Salconi, Saleem, Salemi, Samajdar, Sanchez, Sanchez, Sanchez, Sanchis-Gual, Sanders, Sanuy, Saravanan, Sarin, Sassolas, Satari, Sathyaprakash, Sato, Sato, Sauter, Savage, Sawada, Sawant, Sawant, Sayah, Schaetzl, Scheel, Scheuer, Schiworski, Schmidt, Schmidt, Schnabel, Schneewind, Schofield, Schönbeck, Schulte, Schutz, Schwartz, Scott, Scott, Seglar-Arroyo, Sekiguchi, Sekiguchi, Sellers, Sengupta, Sentenac, Seo, Sequino, Sergeev, Setyawati, Shaffer, Shahriar, Shams, Shao, Sharma, Sharma, Shawhan, Shcheblanov, Shibagaki, Shikauchi, Shimizu, Shimoda, Shimode, Shinkai, Shishido, Shoda, Shoemaker, Shoemaker, ShyamSundar, Sieniawska, Sigg, Singer, Singh, Singh, Singha, Sintes, Sipala, Skliris, Slagmolen, Slaven-Blair,
  Smetana, Smith, Smith, Soldateschi, Somala, Somiya, Son, Soni, Soni, Sordini, Sorrentino, Sorrentino, Sotani, Soulard, Souradeep, Sowell, Spagnuolo, Spencer, Spera, Srinivasan, Srivastava, Srivastava, Staats, Stachie, Steer, Steinhoff, Steinlechner, Steinlechner, Stevenson, Stops, Stover, Strain, Strang, Stratta, Strunk, Sturani, Stuver, Sudhagar, Sudhir, Sugimoto, Suh, Sullivan, Sullivan, Summerscales, Sun, Sun, Sunil, Sur, Suresh, Sutton, Suzuki, Suzuki, Swinkels, Szczepańczyk, Szewczyk, Tacca, Tagoshi, Tait, Takahashi, Takahashi, Takamori, Takano, Takeda, Takeda, Talbot, Talbot, Tanaka, Tanaka, Tanaka, Tanaka, Tanaka, Tanasijczuk, Tanioka, Tanner, Tao, Tao, Martín, Taranto, Tasson, Telada, Tenorio, Terhune, Terkowski, Thirugnanasambandam, Thomas, Thomas, Thomas, Thompson, Thondapu, Thorne, Thrane, Tiwari, Tiwari, Tiwari, Toivonen, Toland, Tolley, Tomaru, Tomigami, Tomura, Tonelli, Torres-Forné, Torrie, Melo, Töyrä, Trapananti, Travasso, Traylor, Trevor, Tringali, Tripathee, Troiano, Trovato, Trozzo, Trudeau, Tsai, Tsai, Tsang, Tsang, Tsao, Tse, Tso, Tsubono, Tsuchida, Tsukada, Tsuna, Tsutsui, Tsuzuki, Turbang, Turconi, Tuyenbayev, Ubhi, Uchikata, Uchiyama, Udall, Ueda, Uehara, Ueno, Ueshima, Unnikrishnan, Uraguchi, Urban, Ushiba, Utina, Vahlbruch, Vajente, Vajpeyi, Valdes, Valentini, Valsan, van Bakel, van Beuzekom, Brand, Broeck, Vander-Hyde, van~der Schaaf, van Heijningen, Vanosky, van Putten, van Remortel, Vardaro, Vargas, Varma, Vasúth, Vecchio, Vedovato, Veitch, Veitch, Venneberg, Venugopalan, Verkindt, Verma, Verma, Veske, Vetrano, Viceré, Vidyant, Viets, Vijaykumar, Villa-Ortega, Vinet, Virtuoso, Vitale, Vo, Vocca, von Reis, von Wrangel, Vorvick, Vyatchanin, Wade, Wade, Wagner, Walet, Walker, Wallace, Wallace, Walsh, Wang, Wang, Wang, Ward, Warner, Was, Washimi, Washington, Watchi, Weaver, Webster, Weinert, Weinstein, Weiss, Weller, Weller, Wellmann, Wen, Weßels, Wette, Whelan, White, Whiting, Whittle, Wilken, Williams, Williams, Williams,
  Williamson, Willis, Willke, Wilson, Winkler, Wipf, Wlodarczyk, Woan, Woehler, Wofford, Wong, Wu, Wu, Wu, Wu, Wysocki, Xiao, Xu, Yamada, Yamamoto, Yamamoto, Yamamoto, Yamamoto, Yamashita, Yamazaki, Yang, Yang, Yang, Yang, Yang, Yap, Yeeles, Yelikar, Ying, Yokogawa, Yokoyama, Yokozawa, Yoo, Yoshioka, Yu, Yu, Yuzurihara, Zadrożny, Zanolin, Zeidler, Zelenova, Zendri, Zevin, Zhan, Zhang, Zhang, Zhang, Zhang, Zhang, Zhao, Zhao, Zhao, Zhao, Zheng, Zhou, Zhou, Zhu, Zhu, Zimmerman, Zlochower, Zucker, \& Zweizig}]{gwtc3}
{The LIGO Scientific Collaboration}, {The Virgo Collaboration}, {The KAGRA Collaboration}, {et~al.} 2021, GWTC-3: Compact Binary Coalescences Observed by LIGO and Virgo During the Second Part of the Third Observing Run,  arXiv, \dodoi{10.48550/ARXIV.2111.03606}

\bibitem[{{The LIGO Scientific Collaboration} {et~al.}(2023){The LIGO Scientific Collaboration}, {the Virgo Collaboration}, {the KAGRA Collaboration}, Abbott, Abe, Acernese, Ackley, Adhicary, Adhikari, Adhikari, Adkins, Adya, Affeldt, Agarwal, Agathos, Aguiar, Aiello, Ain, Ajith, Akutsu, Albanesi, Alfaidi, Alléné, Allocca, Altin, Amato, Anand, Ananyeva, Anderson, Anderson, Ando, Andrade, Andres, Andrés-Carcasona, Andrić, Ansoldi, Antelis, Antier, Apostolatos, Appavuravther, Appert, Apple, Arai, Araya, Araya, Areeda, Arène, Aritomi, Arnaud, Arogeti, Aronson, Asada, Ashton, Aso, Assiduo, de~Souza~Melo, Aston, Astone, Aubin, AultONeal, Babak, Badaracco, Badger, Bae, Bae, Bagnasco, Bai, Baier, Baird, Bajpai, Baka, Ball, Ballardin, Ballmer, Baltus, Banagiri, Banerjee, Bankar, Barayoga, Barish, Barker, Barneo, Barone, Barr, Barsotti, Barsuglia, Barta, Bartlett, Barton, Bartos, Basak, Bassiri, Basti, Bawaj, Bayley, Bazzan, Bécsy, Bedakihale, Beirnaert, Bejger, Belahcene, Bell, Benedetto, Beniwal, Benoit, Bentley, BenYaala, Bera, Berbel, Bergamin, Berger, Bernuzzi, Beroiz, Berry, Bersanetti, Bertolini, Betzwieser, Beveridge, Bhandare, Bhandari, Bhardwaj, Bhatt, Bhattacharjee, Bhaumik, Bianchi, Bilenko, Bilicki, Billingsley, Bini, Birnholtz, Biscans, Bischi, Biscoveanu, Bisht, Biswas, Bitossi, Bizouard, Blackburn, Blair, Blair, Blair, Bobba, Bode, Boër, Bogaert, Boldrini, Bolingbroke, Bonavena, Bondarescu, Bondu, Bonilla, Bonnand, Booker, Bork, Boschi, Bose, Bose, Bossilkov, Boudart, Bouffanais, Bozzi, Bradaschia, Brady, Bramley, Branch, Branchesi, Brau, Breschi, Briant, Briggs, Brillet, Brinkmann, Brockill, Brooks, Brooks, Brown, Brunett, Bruno, Bruntz, Bryant, Bucci, Buchanan, Bulik, Bulten, Buonanno, Burtnyk, Buscicchio, Buskulic, Buy, Byer, Davies, Cabras, Cabrita, Cadonati, Cagnoli, Cahillane, Bustillo, Callaghan, Callister, Calloni, Camp, Canepa, Caneva, Cannavacciuolo, Cannon, Cao, Cao, Capistran, Capocasa, Capote, Carapella, Carbognani, Carlassara, Carlin,
  Carpinelli, Carrillo, Carter, Carullo, Diaz, Casentini, Castaldi, Caudill, Cavaglià, Cavalier, Cavalieri, Cella, Cerdá-Durán, Cesarini, Chaibi, Chakalis, Subrahmanya, Champion, Chan, Chan, Chan, Chan, Chan, Chandra, Chang, Chang, Chanial, Chao, Chapman-Bird, Charlton, Chassande-Mottin, Chatterjee, Chatterjee, Chatterjee, Chaturvedi, Chaty, Chatziioannou, Chen, Chen, Chen, Chen, Chen, Chen, Chen, Chen, Chen, Cheng, Chessa, Cheung, Chia, Chiadini, Chiang, Chiarini, Chierici, Chincarini, Chiofalo, Chiummo, Choudhary, Choudhary, Christensen, Chu, Chu, Chua, Chung, Ciani, Ciecielag, Cieślar, Cifaldi, Ciobanu, Ciolfi, Clara, Clark, Clarke, Clearwater, Clesse, Cleva, Coccia, Codazzo, Cohadon, Cohen, Colleoni, Collette, Colombo, Colpi, Compton, Conti, Cooper, Corban, Corbitt, Cordero-Carrión, Corezzi, Cornish, Corsi, Cortese, Coschizza, Cotesta, Cottingham, Coughlin, Coulon, Countryman, Cousins, Couvares, Coward, Cowart, Coyne, Coyne, Craig, Creighton, Creighton, Criswell, Croquette, Crowder, Cudell, Cullen, Cumming, Cummings, Cuoco, Curyło, Dabadie, Canton, Dall'Osso, Dálya, Dana, D'Angelo, Danilishin, D'Antonio, Danzmann, Darsow-Fromm, Dasgupta, Datrier, Datta, Datta, Dattilo, Dave, Davier, Davis, Davis, Daw, Dax, DeBra, Deenadayalan, Degallaix, Laurentis, Deléglise, Favero, Lillo, Lillo, Dell'Aquila, Pozzo, Matteis, D'Emilio, Demos, Dent, Depasse, Pietri, Rosa, Rossi, DeSalvo, Simone, Dhurandhar, Diab, Díaz, Didio, Dietrich, Fiore, Fronzo, Giorgio, Giovanni, Giovanni, Girolamo, Diksha, Lieto, Michele, Pace, Palma, Renzo, Divakarla, Dmitriev, Doctor, Doleva, Donahue, D'Onofrio, Donovan, Dooley, Dooney, Doravari, Dorosh, Drago, Driggers, Drori, Ducoin, Dunn, Dupletsa, Durante, D'Urso, Duverne, Dwyer, Eassa, Easter, Ebersold, Eckhardt, Eddolls, Edelman, Edo, Edy, Effler, Eguchi, Eichholz, Eikenberry, Eisenmann, Eisenstein, Ejlli, Engelby, Enomoto, Errico, Essick, Estellés, Estevez, Etzel, Evans, Evans, Evstafyeva, Ewing, Ezquiaga, Fabrizi, Faedi, Fafone, Fair,
  Fairhurst, Fan, Farah, Farr, Farr, Favaro, Favata, Fays, Fazio, Feicht, Fejer, Fenyvesi, Ferguson, Fernandez-Galiana, Ferrante, Ferreira, Fidecaro, Figura, Fiori, Fiori, Fishbach, Fisher, Fittipaldi, Fiumara, Flaminio, Floden, Fong, Font, Fornal, Forsyth, Franke, Frasca, Frasconi, Freed, Frei, Freise, Freitas, Frey, Fritschel, Frolov, Fronzé, Fujii, Fujikawa, Fujimoto, Fulda, Fyffe, Gabbard, Gabella, Gadre, Gair, Gais, Galaudage, Gamba, Ganapathy, Ganguly, Gao, Gao, Gaonkar, Garaventa, García-Núñez, García-Quirós, Gardner, Gargiulo, Garufi, Gasbarra, Gateley, Gayathri, Ge, Gemme, Gennai, George, Gerberding, Gergely, Ghonge, Ghosh, Ghosh, Ghosh, Ghosh, Ghosh, Giacoppo, Giaime, Giardina, Gibson, Gier, Giri, Gissi, Gkaitatzis, Glanzer, Gleckl, Godoy, Godwin, Goetz, Goetz, Golomb, Goncharov, González, Gosselin, Gouaty, Gould, Goyal, Grace, Grado, Graham, Granata, Granata, Gras, Grassia, Gray, Gray, Greco, Green, Green, Gretarsson, Gretarsson, Griffith, Griffiths, Griggs, Grignani, Grimaldi, Grimm, Grote, Grunewald, Gruson, Guerra, Guidi, Guimaraes, Gulati, Gulminelli, Gunny, Guo, Guo, Gupta, Gupta, Gupta, Gupta, Gurs, Gustafson, Gutierrez, Guzman, Ha, Hadiputrawan, Haegel, Haino, Halim, Hall, Hamilton, Hammond, Han, Haney, Hanks, Hanna, Hannam, Hannuksela, Hansen, Hanson, Harada, Harder, Haris, Harms, Harry, Harry, Hartwig, Hasegawa, Haskell, Haster, Hathaway, Hattori, Haughian, Hayakawa, Hayama, Hayes, Healy, Heidmann, Heidt, Heintze, Heinze, Heinzel, Heitmann, Hellman, Hello, Helmling-Cornell, Hemming, Hendry, Heng, Hennes, Hennig, Hennig, Henshaw, Hernandez, Vivanco, Heurs, Hewitt, Higginbotham, Hild, Hill, Himemoto, Hines, Hirata, Hirose, Ho, Hochheim, Hofman, Hohmann, Holcomb, Holland, Hollows, Holmes, Holt, Holz, Hong, Hough, Hourihane, Howell, Howell, Hoy, Hoyland, Hreibi, Hsieh, Hsieh, Hsiung, Huang, Huang, Huang, Huang, Huang, Hübner, Huddart, Hughey, Hui, Hui, Husa, Huttner, Huxford, Huynh-Dinh, Hyland, Iandolo, Ide, Idzkowski, Iess, Inayoshi, Inoue,
  Iosif, Irwin, Gupta, Isi, Ito, Itoh, Iyer, JaberianHamedan, Jacqmin, Jacquet, Jadhav, Jadhav, Jain, James, Jan, Jani, Janquart, Janssens, Janthalur, Jaranowski, Jariwala, Jarov, Jaume, Jenkins, Jenner, Jeon, Jia, Jiang, Jin, Johns, Johnston, Johny, Jones, Jones, Jones, Jones, Joshi, Ju, Jung, Jung, Junker, Juste, Kaihotsu, Kajita, Kakizaki, Kalaghatgi, Kalogera, Kamai, Kamiizumi, Kanda, Kandhasamy, Kang, Kanner, Kao, Kapadia, Kapasi, Karat, Karathanasis, Karki, Kashyap, Kasprzack, Kastaun, Kato, Katsanevas, Katsavounidis, Katzman, Kaur, Kawabe, Kawaguchi, Kéfélian, Keitel, Key, Khadka, Khalili, Khan, Khanam, Khazanov, Khetan, Khursheed, Kijbunchoo, Kim, Kim, Kim, Kim, Kim, Kim, Kim, Kimball, Kimura, King, Kinley-Hanlon, Kirchhoff, Kissel, Klimenko, Klinger, Knee, Knust, Kobayashi, Koch, Koehlenbeck, Koekoek, Kohri, Kokeyama, Koley, Kolitsidou, Kolstein, Kondrashov, Kong, Kontos, Korobko, Kossak, Kovalam, Koyama, Kozak, Kozakai, Kranzhoff, Kringel, Krishnendu, Królak, Kuehn, Kuijer, Kulkarni, Kumar, Kumar, Kumar, Kumar, Kumar, Kume, Kuns, Kuromiya, Kuroyanagi, Kuwahara, Kwak, Lacaille, Lagabbe, Laghi, Lalande, Lalleman, Lamberts, Landry, Lane, Lang, Lange, Lantz, Rosa, Lartaux-Vollard, Lasky, Lawrence, Laxen, Lazzarini, Lazzaro, Leaci, Leavey, LeBohec, Lecoeuche, Lee, Lee, Lee, Lee, Legred, Lehmann, Lemaître, Lenti, Leonardi, Leonova, Leroy, Letendre, Levesque, Levin, Leviton, Leyde, Li, Li, Li, Li, Li, Li, Lin, Lin, Lin, Lin, Lin, Lin, Linde, Linker, Littenberg, Liu, Liu, Liu, Llamas, Lo, Lo, London, Longo, Lopez, Portilla, Lorenzini, Loriette, Lormand, Losurdo, Lott, Lough, Lousto, Lovelace, Lowry, Lucaccioni, Lück, Lumaca, Lundgren, Lung, Luo, Lussier, Lynam, Ma'arif, Macas, MacInnis, Macleod, MacMillan, Macquet, Hernandez, Magazzù, Magee, Maggiore, Magnozzi, Mahesh, Majorana, Makarem, Maksimovic, Maliakal, Malik, Man, Mandic, Mangano, Mannix, Mansell, Mansingh, Manske, Mantovani, Mapelli, Marchesoni, Pina, Marion, Mark, Márka, Márka, Markakis,
  Markosyan, Markowitz, Maros, Marquina, Marsat, Martelli, Martin, Martin, Martinez, Martinez, Martinez, Martinovic, Martynov, Marx, Masalehdan, Mason, Masserot, Masso-Reid, Mastrogiovanni, Matas, Mateu-Lucena, Matiushechkina, Mavalvala, McCann, McCarthy, McClelland, McClincy, McCormick, McCuller, McGhee, McGinn, McGuire, McIsaac, McIver, McLeod, McRae, McWilliams, Meacher, Mehmet, Mehta, Meijer, Melatos, Mendell, Menendez-Vazquez, Menoni, Mercer, Mereni, Merfeld, Merilh, Merritt, Merzougui, Messenger, Messick, Meyers, Meylahn, Mhaske, Miani, Miao, Michaloliakos, Michel, Michimura, Middleton, Mihaylov, Miller, Miller, Miller, Millhouse, Mills, Milotti, Minenkov, Mio, Mir, Miravet-Tenés, Mishkin, Mishra, Mishra, Mistry, Mitchell, Mitra, Mitrofanov, Mitselmakher, Mittleman, Miyakawa, Miyo, Miyoki, Mo, Modafferi, Moguel, Mogushi, Mohapatra, Mohite, Molina-Ruiz, Mondal, Mondin, Montani, Moore, Moragues, Moraru, Morawski, More, More, Moreno, Moreno, Mori, Morisaki, Morisue, Moriwaki, Mours, Mow-Lowry, Mozzon, Muciaccia, Mukherjee, Mukherjee, Mukherjee, Mukherjee, Mukund, Mullavey, Munch, Muñiz, Murray, Muusse, Nadji, Nagano, Nagar, Nagar, Nakamura, Nakano, Nakano, Nakayama, Napolano, Nardecchia, Narola, Naticchioni, Nayak, Neil, Neilson, Nelson, Nelson, Nery, Neubauer, Neunzert, Ng, Ng, Nguyen, Nguyen, Nguyen, Quynh, Ni, Ni, Nichols, Nieradka, Nishimoto, Nishizawa, Nissanke, Nitoglia, Niu, Nocera, Norman, North, Notte, Novak, Nozaki, Nurbek, Nuttall, Obayashi, Oberling, O'Brien, O'Dell, Oelker, Oertel, Ogaki, Oganesyan, Oh, Oh, Oh, O'Hanlon, Ohashi, Ohashi, Ohkawa, Ohme, Ohta, Okutani, Oliveri, Olivetto, Oohara, Oram, O'Reilly, Ormiston, Ormsby, Orselli, O'Shaughnessy, O'Shea, Oshino, Ossokine, Osthelder, Otabe, Ottaway, Overmier, Pace, Pagano, Pagano, Pagliaroli, Pai, Pai, Pal, Palamos, Palashov, Palomba, Pan, Panda, Pang, Pannarale, Pant, Panther, Paoletti, Paoli, Paolone, Pappas, Parisi, Park, Parker, Pascucci, Pasqualetti, Passaquieti, Passuello, Patel, Patel,
  Pathak, Patricelli, Patron, Paul, Payne, Pedraza, Pedurand, Pegna, Pegoraro, Pele, Arellano, Penano, Penn, Perego, Pereira, Pereira, Perez, Périgois, Perkins, Perreca, Perriès, Perry, Pesios, Petermann, Pfeiffer, Pham, Pham, Phukon, Phurailatpam, Piccinni, Pichot, Piendibene, Piergiovanni, Pierini, Pierra, Pierro, Pillant, Pillas, Pilo, Pinard, Pineda-Bosque, Pinto, Pinto, Piotrzkowski, Piotrzkowski, Pirello, Pitkin, Placidi, Placidi, Planas, Plastino, Poggiani, Polini, Pong, Ponrathnam, Porter, Posnansky, Poulton, Powell, Pracchia, Pradier, Prajapati, Prasai, Prasanna, Pratten, Principe, Prodi, Prokhorov, Prosposito, Prudenzi, Puecher, Punturo, Puosi, Puppo, Pürrer, Qi, Quartey, Quetschke, Quinonez, Quitzow-James, Raab, Raaijmakers, Radkins, Radulesco, Raffai, Rail, Raja, Rajan, Ramirez, Ramirez, Ramos-Buades, Rana, Rana, Rangnekar, Rapagnani, Ray, Raymond, Raza, Razzano, Read, Regimbau, Rei, Reid, Reid, Reinhard, Reitze, Relton, Renzini, Rettegno, Revenu, Reyes, Reza, Rezac, Rezaei, Ricci, Richards, Richardson, Richardson, Riles, Rinaldi, Robertson, Robertson, Robie, Robinet, Rocchi, Rodriguez, Rolland, Rollins, Romanelli, Romano, Romel, Romero, Romero-Shaw, Romie, Ronchini, Roocke, Rosa, Rose, Rosińska, Ross, Rossello, Rowan, Rowlinson, Roy, Roy, Royzman, Rozza, Ruggi, Ruiz-Rocha, Ryan, Sachdev, Sadecki, Sadiq, Saffarieh, Saha, Saito, Sakai, Sakellariadou, Sakon, Salces-Carcoba, Salconi, Saleem, Salemi, Sallé, Samajdar, Sanchez, Sanchez, Sanchez, Sanchis-Gual, Sanders, Sanuy, Saravanan, Sarin, Sasli, Sassolas, Satari, Sathyaprakash, Sauter, Savage, Savant, Sawada, Sawant, Sayah, Schaetzl, Scheel, Scheuer, Schiworski, Schmidt, Schmidt, Schnabel, Schneewind, Schofield, Schönbeck, Schulte, Schutz, Schwartz, Scott, Scott, Seglar-Arroyo, Sekiguchi, Sellers, Sengupta, Sentenac, Seo, Sequino, Sergeev, Servignat, Setyawati, Shaffer, Shahriar, Shaikh, Shams, Shao, Sharma, Sharma, Shawhan, Shcheblanov, Sheela, Sheridan, Shikano, Shikauchi, Shimizu, Shimode,
  Shinkai, Shishido, Shoda, Shoemaker, Shoemaker, ShyamSundar, Sieniawska, Sigg, Silenzi, Singer, Singh, Singh, Singh, Singha, Sintes, Sipala, Skliris, Slagmolen, Slaven-Blair, Smetana, Smith, Smith, Smith, Soldateschi, Somala, Somiya, Song, Soni, Soni, Sordini, Sorrentino, Sorrentino, Soulard, Souradeep, Spagnuolo, Spencer, Spera, Spinicelli, Srivastava, Srivastava, Stachie, Stachurski, Steer, Steinlechner, Steinlechner, Stergioulas, Stops, Strain, Strang, Stratta, Strong, Strunk, Sturani, Stuver, Suchenek, Sudhagar, Sugimoto, Suh, Sullivan, Summerscales, Sun, Sunil, Sur, Suresh, Sutton, Suzuki, Suzuki, Suzuki, Swinkels, Syx, Szczepańczyk, Szewczyk, Tacca, Tagoshi, Tait, Takahashi, Takahashi, Takano, Takeda, Takeda, Talbot, Talbot, Tamanini, Tanaka, Tanaka, Tanaka, Tanasijczuk, Tanioka, Tanner, Tao, Tao, Tapia, Martín, Taranto, Taruya, Tasson, Tenorio, Terhune, Terkowski, Themann, Thirugnanasambandam, Thomas, Thomas, Thomas, Thompson, Thompson, Thompson, Thondapu, Thorne, Thrane, Tiwari, Tiwari, Tiwari, Toivonen, Tolley, Tomaru, Tomura, Tonelli, Torres-Forné, Torrie, e~Melo, Tournefier, Töyrä, Trapananti, Travasso, Traylor, Trenado, Trevor, Tringali, Tripathee, Troiano, Trovato, Trozzo, Trudeau, Tsai, Tsang, Tsang, Tsao, Tse, Tso, Tsuchida, Tsukada, Tsuna, Tsutsui, Turbang, Turconi, Turski, Tuyenbayev, Ubach, Ubhi, Uchiyama, Udall, Ueda, Uehara, Ueno, Ueshima, Unnikrishnan, Urban, Ushiba, Utina, Vahlbruch, Vaidya, Vajente, Vajpeyi, Valdes, Valentini, Vallero, Valsan, van Bakel, van Beuzekom, van Dael, van~den Brand, Broeck, Vander-Hyde, de~Walle, van Dongen, van Haevermaet, van Heijningen, Vanosky, van Putten, van Ranst, van Remortel, Vardaro, Vargas, Varma, Vasúth, Vecchio, Vedovato, Veitch, Veitch, Venneberg, Venugopalan, Verdier, Verkindt, Verma, Verma, Vermeulen, Veske, Vetrano, Viceré, Vidyant, Viets, Vijaykumar, Villa-Ortega, Vinet, Virtuoso, Vitale, Vocca, von Reis, von Wrangel, Vorvick, Vyatchanin, Wade, Wade, Wagner, Walet, Walker, Wallace, Wallace,
  Wang, Wang, Wang, Ward, Warner, Was, Washimi, Washington, Watada, Watarai, Watchi, Wayt, Weaver, Weaving, Webster, Weinert, Weinstein, Weiss, Weller, Weller, Wellmann, Wen, Weßels, Wette, Whelan, White, Whiting, Whittle, Wilk, Wilken, Williams, Williams, Williams, Williamson, Willis, Willke, Wipf, Woan, Woehler, Wofford, Wojtowicz, Wong, Wong, Wright, Wu, Wu, Wu, Wysocki, Xiao, Yadav, Yamada, Yamamoto, Yamamoto, Yamamoto, Yamashita, Yamazaki, Yang, Yang, Yang, Yang, Yang, Yang, Yap, Yeeles, Yeh, Yelikar, Yokoyama, Yokozawa, Yoo, Yoshioka, Yu, Yu, Yuzurihara, Zadrożny, Zanolin, Zeidler, Zelenova, Zendri, Zevin, Zhan, Zhang, Zhang, Zhang, Zhang, Zhang, Zhang, Zhao, Zhao, Zhao, Zhao, Zheng, Zhou, Zhu, Zhu, Zimmerman, Zucker, \& Zweizig}]{o3_lensing}
{The LIGO Scientific Collaboration}, {the Virgo Collaboration}, {the KAGRA Collaboration}, {et~al.} 2023, Search for gravitational-lensing signatures in the full third observing run of the LIGO-Virgo network.
\newblock \doarXiv{2304.08393}

\bibitem[{Veitch \& Vecchio(2010)}]{veitch_vechhio_likelihood}
Veitch, J., \& Vecchio, A. 2010, Phys. Rev. D, 81, 062003, \dodoi{10.1103/PhysRevD.81.062003}

\bibitem[{Wang {et~al.}(1996)Wang, Stebbins, \& Turner}]{wang_lensing}
Wang, Y., Stebbins, A., \& Turner, E.~L. 1996, Phys. Rev. Lett., 77, 2875, \dodoi{10.1103/PhysRevLett.77.2875}

\bibitem[{Wempe {et~al.}(2022)Wempe, Koopmans, Wierda, Hannuksela, \& Broeck}]{Wempe:2022zlk}
Wempe, E., Koopmans, L. V.~E., Wierda, A. R. A.~C., Hannuksela, O.~A., \& Broeck, C. v.~d. 2022, {A lensing multi-messenger channel: Combining LIGO-Virgo-Kagra lensed gravitational-wave measurements with Euclid observations}.
\newblock \doarXiv{2204.08732}

\bibitem[{Wierda {et~al.}(2021)Wierda, Wempe, Hannuksela, Koopmans, \& Van Den~Broeck}]{wierda_rgal}
Wierda, A. R. A.~C., Wempe, E., Hannuksela, O.~A., Koopmans, L. e. V.~E., \& Van Den~Broeck, C. 2021, Astrophys. J., 921, 154, \dodoi{10.3847/1538-4357/ac1bb4}

\bibitem[{Williams {et~al.}(2021)Williams, Veitch, \& Messenger}]{nessai}
Williams, M.~J., Veitch, J., \& Messenger, C. 2021, Physical Review D, 103, \dodoi{10.1103/physrevd.103.103006}

\bibitem[{Wright \& Hendry(2022)}]{gravelamps_paper}
Wright, M., \& Hendry, M. 2022, The Astrophysical Journal, 935, 68, \dodoi{10.3847/1538-4357/ac7ec2}

\bibitem[{Wright {et~al.}(2022)Wright, Liu, Seo, \& Wong}]{gravelamps_software}
Wright, M., Liu, A., Seo, E., \& Wong, I. C.~F. 2022, {\textsc{Gravelamps}: Gravitational Wave Lensing Mass Profile Model Selection}, \url{https://github.com/mick-wright/Gravelamps}

\bibitem[{Xu {et~al.}(2022)Xu, Ezquiaga, \& Holz}]{xu_rate}
Xu, F., Ezquiaga, J.~M., \& Holz, D.~E. 2022, The Astrophysical Journal, 929, 9, \dodoi{10.3847/1538-4357/ac58f8}

\end{thebibliography}

\end{document}